\documentstyle[12pt,aas2pp4]{article}

\lefthead{Muno, Morgan, \& Remillard}
\righthead{QPOs and Spectral States in GRS 1915+105}

\begin{document}

\title{QPOs and Spectral States in GRS 1915+105}
\author{Michael P. Muno, Edward H. Morgan, and Ronald A. Remillard}
\affil{Center for Space Research}
\authoraddr{77 Massachusetts Ave., Cambridge, MA 02139}
\authoremail{muno@mit.edu, ehm@space.mit.edu, rr@space.mit.edu}
\begin{abstract}

We present results from the analysis of X-ray energy spectra and quasi-
periodic oscillations (QPOs) from a set of observations which samples a 
broad range of time variability in GRS 1915+105. We first demonstrate that the 
frequency and integrated amplitude of a 0.5-10 Hz QPO is correlated 
with the apparent temperature 
of the accretion disk for the majority of observations.
We then show that the behavior of GRS 1915+105 exhibits two distinct
modes of accretion. In the first mode, the QPO is present between 
0.5--10 Hz, variability in the source luminosity is dominated by the power law 
component. In the second mode, the QPO is absent, the changes in 
the luminosity are dominated by thermal emission from the accretion disk. 
We find that the color radius and temperature of
the inner accretion disk are empirically related by 
$R_{col} \propto T_{col}^{-2} + const$. We discuss these results in terms of 
ongoing efforts to explain the origin of both the QPOs and the hard
X-ray component in the spectrum of GRS 1915+105.

\end{abstract}


\section{Introduction}

GRS 1915+105 is a transient X-ray source that has been extremely active during 
the six years since it was discovered in 1992 with the
WATCH instrument on {\it Granat} (\cite{ct92}).
It is located behind the Sagittarius arm of the Milky Way at an estimated 
distance of 12.5 kpc (\cite{mr94}), where extinction from interstellar 
dust limits optical/IR studies 
to wavelengths greater than 1 $\mu$m (\cite{mir94}). 
No measurement has yet been made of a binary mass function or orbital
period.
GRS 1915+105 is one of several galactic X-ray sources
observed to produce superluminal radio jets (\cite{mr94}, \cite{fen99}). 
One of these sources, GRO J1655-40 (\cite{zha94}), has been observed 
optically to be a binary system containing normal F star and a 7 M$_\odot$ 
compact
object presumed to be a black hole (\cite{ob97}). Since the spectral  
properties of GRS 1915+105 are similar to those of GRO J1655-40
(\cite{gro98}, \cite{rem98}), and the luminosity of GRS 1915+105 in outburst
is $5 \times 10^{39}$ ergs ${\rm s}^{-1}$ (25 times the Eddington 
luminosity of a neutron star; \cite{gmr97}), it is 
thought that GRS 1915+105 also 
contains a black hole.

The X-ray spectrum of GRS 1915+105 is typical of a black hole 
candidate, and spectral models require at least two emission components. 
The energy spectrum
below 10 keV is dominated by emission which appears to be thermal in 
origin. This is usually modeled with a multi-temperature blackbody 
representing an optically thick, geometrically thin accretion disk 
(\cite{mit84}). 
The spectrum above 10 keV can be modeled with a power law function, 
and is thought to originate from inverse-Compton scattering (\cite{st80}). 
In the case of GRS 1915+105, this power law component
is sometimes seen at energies up to 600 keV (\cite{gro98}).
The physical origin and spatial distributions of the Comptonizing 
electrons is unknown. It was thought that the electrons were part
of an optically thin corona above the plane of the accretion disk, but
recent numerical simulations indicate that a self-consistent planar corona
cannot produce the spectra seen in the low state of black hole binaries
such as Cygnus X-1 (\cite{dov98}).  
Various recent models have suggested that the optically thick 
flow may give way to an optically thin flow close to 
the black hole in a manner which would produce relativistic electrons
 (\cite{chen95}; \cite{ct95}; \cite{ll98}; \cite{tz98}; \cite{dov98}). 
These Comptonizing electrons could either be extremely hot,
or they could be part of a bulk flow of matter
streaming toward the black hole. It is also possible that the 
relativistic electrons are contained in a bulk outflow or a jet.

The X-ray variability of GRS 1915+105 is spectacular 
(see Figure \ref{lcurves}).
Observations with BATSE on the {\it Compton Gamma Ray Observatory} 
(\cite{har94}) and with SIGMA on
{\it Granat} (\cite{fig94}) have revealed that GRS 1915+105 is highly 
variable in the hard X-rays. When the 
{\it Rossi X-Ray Timing Explorer (RXTE)} began observations in
1996, variations of as much as 3 Crab were observed on time scales 
from seconds to days (\cite{gmr97}). 

The variations of the X-ray intensity and spectrum on time scales of hundreds 
of seconds have invited several interpretations. Most models involve a 
thermal-viscous instability in an accretion disk (\cite{bel97a}, 1997b) 
and some take into account the dissipation of accretion energy in 
a hot corona (\cite{tcs97}). 
Moreover, some of these cycles of variability in X-rays
have been strongly linked to non-thermal flares in the 
infrared (\cite{sam96}, \cite{ef97}, \cite{eik98}) and in the radio 
(\cite{mir98}, \cite{fp98}). 
These studies have produced the first observational evidence that directly 
links the formation of jets to instabilities in the accretion disk.

Quasi-periodic oscillations (QPOs) seen in power density spectra (PDS) of GRS 
1915+105 are another area of interesting research 
(\cite{mrg97}; \cite{cts97}). One QPO with a centroid frequency of 67 Hz
appears occasionally, and is likely 
caused by one of several effects due to general relativity in the inner
accretion disk. Common lower frequency QPOs (0.001 -- 10 Hz) 
are broadened in frequency by a random walk in phase, and
exhibit phase lags of a few percent at $\sim 10$ keV relative to 
2 keV. The QPO amplitude increases with photon energy, indicating 
that these QPOs are associated with the hard X-ray power 
law component. Further studies by Markwardt, Swank, \& Taam (1999) 
and Trudolyubov, Churazov, \& Gilfanov (1999) have 
shown that the {\it frequency} 
of a spectrally hard QPO between 0.5-10 Hz is positively correlated with the 
thermal flux from the disk. Thus this QPO appears to be linked {\it both} 
to the accretion disk and the population of Compton scattering electrons.

QPOs are therefore a promising means of probing the relationship between the 
hard and soft components in the spectra of accreting black holes. 
In this paper we examine the relationship between the properties of the
0.5--10 Hz QPO and X-ray spectral parameters from
27 observations, which sample both steady states and 
repetitive patterns of variability. We find that
the frequency and fractional normalization of the QPO are best correlated
with the temperature of the 
inner accretion disk.
We then show that the source has two distinct tracks of spectral evolution, 
which can be distinguished by the presence or absence of the intermediate
frequency (0.5--10 Hz) QPOs. (Note that these QPOs are to be distinguished
from the 67 Hz QPO and from the occasional QPOs seen at lower frequency 
(0.05--0.2 Hz)). 
We find a relationship between the radius and temperature of the
inner disk which spans these two spectral states, and examine the relationship
between the photon
index of and flux from the power law and the parameters of the inner disk. 
We conclude that the 0.5--10 Hz QPO is crucial in understanding the origin 
of the power law component and the variable X-ray emission from GRS 1915+105.

\section{Data Selection and Analysis}

There have been over 300 observations of GRS 1915$+$105  
by the Proportional Counter Array (PCA) (\cite{jah96}) and the High-Energy
X-ray Timing Experiment (HEXTE) (\cite{roth98}) on the {\it Rossi X-ray Timing
Explorer (RXTE)}.  
In this paper, we report on our studies of selected observations (see 
Table \ref{obs}). These observations represent 15\% of the time spent 
observing GRS 1915+105 with the PCA and HEXTE through December 1998.
Although these observations are not representative of the amount
of time GRS 1915+105 spends in any one state, they cover all types 
of variability seen to date.

\placetable{obs}

\begin{figure} 
\plotone{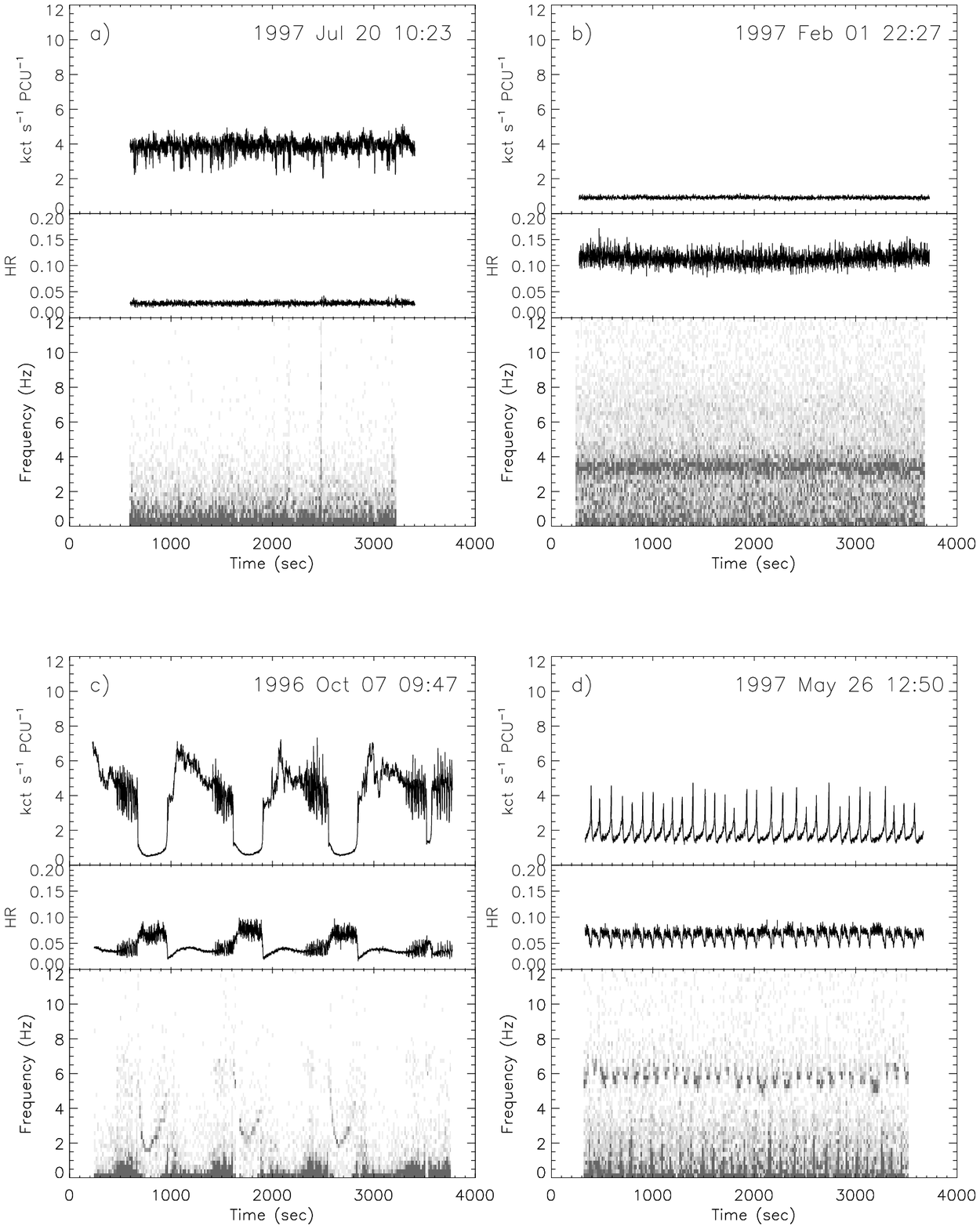}
\caption{Light curves, PCA HRs, and dynamic PDS of GRS 1915+105 from the PCA 
on {\it RXTE}. Light curves (top panels) are the count rate per second per
PCU. We have not corrected
for PCA dead time, which introduces an error of 10\% per 5000 counts/s/PCU.
The PCA HRs (middle panels) are the count rate at 13-25 keV relative 
to the rate at 2-13 keV. The dynamic PDS (bottom panels) are power density 
spectra computed
every sixteen seconds and rebinned at 0.25 Hz, plotted with time on the 
horizontal axis, frequency on the vertical axis, and the linear Leahy power 
density represented by the grey-scale from white (0) to 
black ($>$ 50). A QPO in the PDS appears as a dark horizontal band.}
\label{lcurves}
\end{figure}

Figure \ref{lcurves} displays
X-ray light curves, PCA hardness ratios (HRs), and ``dynamic power spectra'' 
(dynamic PDS) which are demonstrative of the types of variability in 
the observations listed in Table \ref{obs}.  The light curves represent the
count rate in the PCA band (2--30 KeV) per proportional 
counting units (PCU). The PCA HRs are the count rate at 13-25 keV relative 
to the rate at 2-13 keV. The dynamic PDS are power density spectra computed
every sixteen seconds and rebinned at 0.25 Hz, plotted with time on the 
horizontal axis, frequency on the vertical axis, and the linear Leahy power 
density represented by the grey-scale from white (0) to 
black ($>$ 50). 

Figure \ref{lcurves}a characterizes the observations on 1996 May 5, 
1997 July 20, 1997 November 17, 1998 February 3, and 1998 February 14. 
These observations have steady count rates and do not exhibit the narrow 
QPO between 0.5--10 Hz, although low frequency or broad 
QPOs are present in some of these observations (see Table \ref{obs}). 
We find that HR $<$ 0.05 for this set of observations, so we refer
to this first set of observations as ``soft-steady''. 

Figure \ref{lcurves}b is representative of the observations during three
long time intervals: 1996 July 11 
through 1996 August 3; 1996 November 28 through 1997 March 26; 
and 1997 October 9 through 25. All of these observations have a steady count 
rate and a strong 0.5--10 Hz QPO. Their spectra are harder 
(0.08 $<$ HR $<$ 0.15) than the soft-steady group, and Morgan, 
Remillard, \& Greiner (1997)\markcite{mrg97} label these as ``low-hard
states''. 
Chen, Taam, \& Swank (1997)\markcite{cts97} discuss the hardness ratio, 
count rate, and PDS of the observations in the first time interval, and
Trudolyubov, Churazov \& Gilfanov (1999) discuss energy spectra
and PDS of the observations from the second time interval. 
The radio flux at 8.3 GHz (or 15.2 GHz for the first interval; see 
Table \ref{obs}) distinguishes two subsets in these 
intervals: the values listed in Table \ref{obs} are greater than 35 mJy 
during the {\it RXTE} observations we used from the 
first and third intervals, but less than 
15 mJy during the observations from second interval
(there is a radio flare to 110 mJy on 1996 December 6, but there 
was no coincident {\it RXTE} observation on that date). 
Therefore we refer to the {\it RXTE} observations we used from these 
intervals as the ``radio-loud hard-steady'' states and ``the radio-quiet 
hard-steady'' states respectively. The radio-loud observations are
given particular attention later in this paper, as the parameters we 
derive from spectral fits to these observations are difficult to interpret. 

The remainder of the observations exhibit a wide range of variability.
Figure \ref{lcurves}c is taken from the observation on 1996 October 7.
The dips in this observation are spectrally hard and contain a 0.5--10 Hz 
QPO, while the brighter portions are soft and void of this QPO. Theoretical 
models of thermal-viscous instabilities in an optically thick accretion
disk have been used by Belloni et al. (1997a) \markcite{bel97a} to 
explain this series of dips; based upon spectral analyses they conclude that 
the inner disk empties and re-fills over the course of each dip.

The light curve in Figure \ref{lcurves}d is from the observation on 1997 May 
26. The time-series exhibits a QPO
throughout the low portion of the light curve in addition to the 
large quasi-periodic ``ringing'' flares every $\sim$ 120 s. 
Taam, Chen, \& Swank (1997) \markcite{tcs97} have used numerical simulations 
of an unstable accretion disk which dissipates energy into a hot, optically 
thin corona to explain many features of these rapid bursts.

The observations on 1997 August 14, 1997 September 9, and 
1997 October 30 are similar to the light curve in Figure \ref{lcurves}e.
The longer dips are hard, and contain a QPO; the other features lack the 
0.5--10 Hz QPO, and are generally soft. These observations are distinct 
from those represented by Figure \ref{lcurves}c in that they display a 
large ``spike'' at the end of the long dips, which is 
followed by a spectrally soft dip. 
Observations of this type exhibit infrared and radio  
flares which follow the X-ray dip-spike cycle (see references in introduction).
The hard X-ray dips can also 
be explained by the inner disk emptying and re-filling. The observation on 
1997 September 9 has been analyzed by Markwardt, Swank, \& Taam (1999)
\markcite{mst98} as well, in a manner similar
to this paper.

The light curve in Figure \ref{lcurves}f is
from the observation on 1997 September 16. The dip/flare
cycles have properties which are the reverse 
of cases 1e and 1c, in that the lowest dips in the observation
have soft spectra and lack a 0.5--10 Hz QPO, while the brighter portions
are spectrally harder and contain a QPO. Spectral analyses indicate that 
these patterns are not consistent with the disk instability model of Belloni 
et al. (1997b).

High-luminosity soft states are presented in Figure \ref{lcurves}g (from
1997 August 13) and Figure \ref{lcurves}h (representative of 1997 August 19 
and 1997 December 22). None of these observations contain a narrow 0.5-10
Hz QPO. The observation
in Figure \ref{lcurves}g is presented in Remillard et al. (1998) in
a discussion of the similarities between GRS 1915$+$105 and GRO J1655-40.

Figure \ref{lcurves}i illustrates a moderately soft and
bright interval from the observation on (1997 September 18). During the
course of the observation, the count rate increases 
and the spectrum softens on time scales longer than the {\it RXTE} orbit 
for which data is displayed. 
This observation exhibits the 0.5--10 Hz QPO, which is particularly 
strong at lower count 
rates and higher HR. This observation also presented some
problems with spectral fits, which we will address later
in this paper. 

Finally, the observation in Figure
\ref{lcurves}j displays another type of X-ray ringing (1997 May 18). It 
exhibits a series of soft flares ($\sim$ 100 seconds long) that 
recur with increasingly longer time intervals and lower amplitudes, until 
the series terminates in a long (1000 s), hard minimum in the 
count rate. Thereafter, the cycle begins again. The 0.5--10 Hz 
QPO is present throughout the observation.
 
In order to create energy spectra and PDS with good statistics while avoiding 
the intrinsic changes due to the chronic
variability of GRS 1915+105, we separate the observations into three 
categories that we analyze differently. If the standard deviation in the 
PCA count rate over the full energy band width (effectively 2--30 keV) in 1 s 
bins is less than 15\% of 
the mean count rate during every 96 min {\it RXTE} orbit of an observation, 
we collect a single energy 
spectrum and PDS for each  orbit. 
The resulting interval of on-source exposure time outside of the 
South Atlantic 
Anomaly is generally $\sim 3000$ s. If the count rate changes slowly
from orbit to orbit, but during a single orbit the variability is less than 
15\%, we collect energy spectra and PDS every 512 s. If the count rate in
each orbit varies by more than 15\%, we create energy spectra and PDS every 
32 s to track the changes in the light curve. Finally, if the count rate varies
by more than 15\% during 32 s, the time segment is removed from our analysis.
Subsequent to these data selections, each spectrum and PDS is analyzed 
identically.

\subsection{Energy Spectra}

We perform fits to the energy spectra from both 
the PCA (Table \ref{fits}) and HEXTE (Table \ref{hexte}). We use 
the standard background subtraction procedures for each instrument, 
and apply these to 128 channel spectra from the Standard 2 mode
of the PCA and 64 channel spectra from the Archive mode of HEXTE.
All spectra are first fit in the PCA band alone. 
We then fit the combined PCA/HEXTE spectra for the steady observations 
only, because the count rate from GRS 1915+105 in the HEXTE band is too 
low to analyze with good statistics 
on 32 s time scales. We also are unable to analyze HEXTE spectra from 
the steady observations 
on 1996 May 05, July 11, July 26, or August 03, because the HEXTE clusters 
were not rocking between source and background positions, and accurate 
background estimates are not available.
This leaves 12 observations for which we provide the results of combined
PCA/HEXTE spectral fits (Table \ref{hexte}). 

In order to investigate systematic errors in the PCA and HEXTE response 
matrices (from 1997 October 2 and 1997 March 20 respectively), we have
analyzed spectra from the Crab Nebula before modeling more complex spectra
from GRS 1915+105. We fit the Crab spectrum with a model consisting of
a power law with photo-electric absorption. There is sufficient curvature 
in the Crab spectrum to prevent an adequate fit over the complete bandwidth
of the PCA and HEXTE. However, the power law fit to each instrument is
good, and we use the results to identify persistent 
local features in the residuals of an individual detector unit. 
To lessen the statistical weight of such features, we add 1\% systematic 
errors to the 
PHA bins in both the PCA and HEXTE. In spectra from the PCA, there 
are larger systematic deviations in the residuals below 2.5 keV and above 
25 keV,
which a leads us to limit the energy range for PCA analysis to 
2.5--25 keV. Moreover, the spectral fits from PCUs 2 and 3 have 
systematically larger residuals between 5 and 7 keV than do fits from 
the other PCUs, which leads us to use only data from PCUs 0, 1, and 4.
We use both HEXTE clusters, and we find systematic deviation in HEXTE fits 
to the crab spectrum below 15
keV, which leads us to limit our analysis of HEXTE spectra to energies
greater than 15 keV. The upper limit to fits to HEXTE spectra is determined
by the energy above which the source is no longer detectable, which 
occurs between 30--170 keV depending on the observation. 

We have applied many models to our spectral analysis of GRS 1915+105, and we
find that 22 out of 27 of the PCA spectra are best fit
by the standard disk black body and power law component model.
We also needed a Gaussian emission line (with a fixed FWHM of 1.0 
keV) to measure iron emission between 6 and 7 keV. 
When fitting all of our spectra, the column density is 
fixed to $N_H = 6.0\times10^{22}$ cm$^{-2}$ of H, which was chosen by 
allowing the column density to float in several spectral fits and taking
the average of the resulting values. Finally, we add a fixed multiplicative 
constant normalization on each PCU and each HEXTE cluster (when applicable)
to account for differences in the effective areas of the detectors.
This is the same model that has been used for GRS 1915+105 spectra by
Belloni et al. (1997) and Taam, Chen, and Swank (1997).   

There are several systematic issues that cause us to use caution when 
interpreting
the absolute values of the parameters quoted throughout this paper. 
The multi-temperature disk model does not take into account
electron scattering (\cite{ebi93}, \cite{st95}) and 
general relativistic effects at the inner disk (\cite{zcc97}), which modify
the emergent spectrum. It is necessary to correct the observed
model parameters for these effects in order to obtain estimates of
the physical parameters of the disk. The accuracy of such corrections is 
largely uncertain. Moreover, the black hole mass 
is not known, and the temperature and radius of the inner disk appear
wildly variable in GRS 1915+105. In light of these problems we refrain
from applying any of these corrections until 
the discussion in Section 4.

The free parameters in our model for the disk emission  
are the normalization on the disk black body component ($N_{bb}$) and the 
color temperature of the disk at the inner radius 
($T_{col}$). The characteristic radius of the 
inner disk  ($R_{col} = D_{10 kpc}\sqrt{N_{bb}/cos \theta}$ km) is derived 
from the normalization of the disk black body ($N_{bb}$), assuming a 
distance of 12.5 kpc ($D_{10 kpc}=1.25$) and 
an inclination angle ($\theta$) equal to that of the radio jets, 
$70^{\circ}$ (\cite{mr94}). The total flux from the disk 
is then $F_{bb} = 1.08\times 10^{-11} N_{bb} \sigma T_{col}^4$
ergs$^{-1}$cm$^{-2}$s, where $\sigma$ is the Stephan-Boltzmann constant. 
Our model parameters for the power law component are the 
photon index of the power law component ($\Gamma$) and 
the flux at 1 keV from the hard component ($N_{\Gamma}$).  
The flux from the power law ($F_{pl}$) is calculated by integrating 
$1.60 \times 10^{-9} N_{\Gamma} E^{-\Gamma+1}$ from 1 to 25 keV, where $E$ 
is energy in keV. 
Finally, our model provides the centroid ($E_{gauss}$) and 
normalization ($N_{gauss}$) of the Gaussian. 

\placetable{fits}

The parameters, fluxes, and reduced chi-squared values for the PCA spectral
fits are presented in Table \ref{fits}. In 21 of the 27 cases the 
standard spectral model provides the best fit with a reduced 
chi-squared $< 2.0$. The observation of 1997 September 18 on average 
yields a higher value ($\chi^2_{\nu} = 3.18$), but we find no 
better alternative model. The remaining
five spectra (from the soft-steady observations, identified
by the use of $E_c$ in Table \ref{fits}) are not 
consistent with our standard model. The reduced chi-squared values for these
observations were initially in 
the range of 3 to 35, and we were forced to adopt an alternative model. 
After exploring many options,  
and found that the hard ($> 10$ keV) emission in these observations falls 
off as an exponential in energy rather than a power law. 
Although several more complicated physical models (e.g. the thermal
Comptonization model of \cite{tit93}) 
also fit these spectra well, we are unable to constrain 
the extra parameters used in these spectral models with our data. 
For these five cases we therefore replace the power law 
with an exponential photon spectrum, $N_{\Gamma}\exp(-E/E_c)$, 
where $E$ is the photon energy, $E_c$ is the cut-off
energy of the exponential, and $N_{\Gamma}$ is flux at 1keV. 
The flux from the exponential ($F_{exp}$) is calculated 
by integrating over the energy range 1 to 
25 keV. The final reduced chi-squared values for all five of these 
remaining observations are in the range of 0.5--2.6, and the corresponding
disk parameters (e.g. $T_{col}$, see Table \ref{fits}) are very similar to the 
range of results for observations in which the hard component is a power law.

We must further note that fits with a disk black body and a power law in 6 
of the observations (viz. the 5 observations
in the radio-loud hard-steady state, plus many segments of 1997 September 18) 
yield very small 
values for the color radius ($<$ 10 km) and high values for the effective 
temperature of the inner disk (3-5 keV).
Similar episodes have also been observed in GRS 1124-68 (Nova Muscae; 
\cite{ebi94}), GRO J1655-40 (\cite{sob99a}), and XTE J1550-524 
(\cite{sob99b}) in which the disk spectrum appears very hot
with a normalization that implies a small characteristic radius.
These observations are discussed further in Section 4.

\placetable{hexte}

The results for the combined PCA/HEXTE fits are given in Table \ref{hexte}.
The high-energy upper limits for these fits ($E_{max}$) are given in column 2.
As noted above, joint PCA/HEXTE fits are statistically meaningful 
only for long exposures during the steady states in GRS 1915+105. 
For the soft-steady and
radio-loud, hard-steady observations, the results from the PCA/HEXTE fits 
are consistent with those derived from the PCA alone. However, for the
radio-quiet, hard-steady observations, we find
that the addition of HEXTE data to our analysis requires that we add to our
standard model either a high-energy cut-off to the power law (see Table
\ref{hexte}) or a 
component representing reflection from un-ionized matter (not shown) in order
to obtain reduced chi-squared values near 1. The reflection model adds a 
single free parameter, the relative amount of hard flux which is reflected.
Assuming that the inclination of the disk is $70^{\circ}$ to the line
of sight and that metals in the disk are of Solar abundance, we sometimes
find values of the reflection parameter $>$ 1.0, which is not physically
possible. Consequently, we have chosen to use the cut-off power law in 
characterizing the hard-steady spectra.
The addition of a cut-off to the power law introduces a systematic decrease of
$\Gamma \sim 0.2$ for the power law, but leaves the disk spectral parameters 
more or less unchanged.

Our results with a cut-off power law are consistent with those of
Trudolyubov, Churazov, \& Gilfanov (1999), who noticed a cut-off 
in the power law became detectable when GRS 1915+105 was at low 
luminosity throughout 
1996 November to 1997 March. The presence of a cut-off in the 
power law would supplement the results of Grove et al. (1998),
who demonstrated that GRS 1915+105 was one of a class of black hole
candidates which exhibit power law spectra with slope 2.7 which can extend 
to 600 keV without a cut-off. If our modification of the model 
to include a cut-off energy is correct, then Table \ref{hexte} would
imply that two types of cut-off power law (with $E_c \sim 3.5$ and 100 keV)
are exhibited by GRS 1915+105. Clearly, more observations in the
100--600 keV range are needed in order to determine whether the 
power law in GRS 1915+105 evolves as a function of time.

\subsection{Power Density Spectra}

To create power density spectra we use data which effectively covers 
2--30 keV with a time resolution of 122 $\mu$s. 
For data segments longer than 32 s, the light curve is divided into 
256 s intervals, and a PDS is created for each interval. The PDS
are then averaged for each segment, weighted by the total 
counts, and the results are logarithmically rebinned and subtracted
for dead-time-corrected Poisson noise (\cite{mrg97}; \cite{zha96}).
The shape of the PDS are as diverse as those in Morgan,
Remillard, \& Greiner (1997) and Chen, Swank, \& Taam (1997).
When the energy spectra are analyzed in 32 s intervals, PDS are created 
for the identical intervals. These spectra are linearly rebinned into 
0.25 Hz bins, but otherwise treated as above.

We search for a QPO peak in the PDS by fitting frequency intervals between 
0.5--12 Hz with a Lorentzian profile on top of a power-law background 
continuum. Only features with a $Q >$ 3 are considered as candidate QPOs. 
In addition we varied this range when low-frequency noise obviously
dominated a portion of the PDS; see Figure \ref{lcurves}d, for instance.
To compensate for systematic difficulties in fitting the QPO profiles, 
we estimate the significance of the QPO to be the ratio of its amplitude 
to the average statistical uncertainty over the QPO width, divided by the
square root of the reduced-chi squared value from the fit. We estimate the 
statistical errors on the parameters of the QPO to be the values of the 
covariance 
matrix from the least chi-squares minimization routine. The average 
uncertainty is $\pm 0.04$ Hz in frequency, $\pm 0.04$ Hz in width, 
and $\pm 0.02$ in amplitude (which is expressed as an RMS deviation
divided by the mean count rate).

In order to investigate changes in the energy spectrum which are correlated 
with the properties of the 0.5--10 Hz QPO, we separate our database of 
PCA spectral and QPO 
parameters into three groups based on the strength of the observed QPOs: 
(1) definite QPOs, for which the best candidate QPO between 
0.5--12 Hz has a significance greater than 6.0; (2) no QPO, for which the 
best QPO candidate either has a significance less than 2.0
or a significance less than 3 with an amplitude less than 2\%; 
and (3) ambiguous cases not selected with the previous criteria. 
The third category is ignored in this paper, as the results are inconclusive or
of poor quality. 

The values for the significances of the QPOs used to define these selection 
criteria are chosen in order to minimize the number 
of false assignments among the large number of trial fits. 
The joint condition on the amplitude and significance of candidate features in 
category 2 is chosen because many of the PDS
contain broad QPOs ($Q <$ 3), knees, and curvature in the background 
continuum that affect the distribution of the amplitudes of marginally 
significant features.

Out of the 250 ks of exposure time denoted by Table \ref{obs},  
54\% yield QPO detections, 15\% show no QPO, and 31\% are dropped 
from further study (including those times when the standard deviation in the
count rate is above 15\%, so that no QPO search or spectral 
analysis is performed). The amplitude of the definite 
QPOs (category 1) is $9 \pm 1$\% for the PDS covering an {\it RXTE} orbit, 
and $7 \pm 3$\% for the PDS from 32 s and 512 s intervals. The amplitude 
of the best candidate feature in the no QPO case (category 2) is 
$0.6 \pm 0.3$\% for PDS 
covering and orbit, and $0.8 \pm 0.8$\% for the PDS from 32 s and 
512 s intervals. The distributions of the ``with QPO'' and ``without QPO''
groups are therefore statistically well separated.

We must also note that because the 32 s exposures have 
limited statistics, we expect that there will be some errors in assigning
these data to each of the three groups. The
dynamic PDS in Figure 1 demonstrates that the 0.5--10 Hz QPO which dominates 
the PDS in this frequency range is generally persistent, and varies slowly 
in frequency so long as the energy spectrum varies slowly. 
However, in rare instances the QPO disappears in a single 32 s time 
interval among a series of intervals that otherwise exhibit a persistent
QPO. In addition, visual inspections of QPO fits reveal occasions when 
a QPO is found in low frequency ($<$ 3 Hz) noise during a single time 
interval when there is no evidence in the dynamic PDS of a persistent QPO.  
Of the $\sim 1000$ points plotted in the figures below, we estimate that 
50 points may have been mis-categorized in one of these two manners.
Finally, we must emphasize that we have restricted our interest to 
features with $Q > 3$ and to the frequency range of 0.5--10 Hz, which 
implicitly presumes that occasional QPOs at both low frequencies 
($\sim 0.1 Hz$) and high frequencies (e.g. 67 Hz) are different X-ray 
oscillation phenomena from the QPOs we examine in this paper.

\section{Results}

In Figures \ref{qpofreq}--\ref{alphat} we present the results of our 
analysis of PCA energy spectral and 0.5--10 Hz QPO parameters. 
The size of the symbol in each plot
corresponds to the length of the time interval from which the data point was
taken: the large squares correspond to data points taken from 
entire {\it RXTE} orbits, the medium-sized asterisks correspond to 
data points from 512 s intervals, 
and the small points correspond to data from 32 s intervals. 
Most of our conclusions use data in the aggregate, but for the reader who
wishes to examine how each type of variability (as explained in Section 2 
and illustrated in Figure \ref{lcurves}) relates to the figures, we plot
each type of variability with a separate color. In our scheme, the most red 
symbols indicate variability types with the softest spectra, and the blue 
those with the hardest spectra. The key for 
the large symbols is as follows: the blue squares are radio-quiet hard-steady 
observations, the green squares are radio-loud hard-steady observations,
and the red squares are soft-steady observations (which were fit with 
an exponential rather than a power law).
Only one observation (1997 September 18; Figure \ref{lcurves}i) was
analyzed at 512 s intervals; this observation is indicated by medium-sized
green asterisks.  
The color code for the 32 s points is as follows: red points correspond to 
the observations similar to Figures \ref{lcurves}g and h; orange points
to Figure \ref{lcurves}f; yellow-green to Figure \ref{lcurves}c;
green points to Figure  \ref{lcurves}d; 
blue to Figure \ref{lcurves}e; and purple to Figure \ref{lcurves}j.
Note that few red symbols appear in Figures \ref{qpofreq}--\ref{qpoq}, as
there is no persistent 0.5--10 Hz QPO evident in these soft observations.

\subsection{The Relationship Between QPO and Spectral Parameters}

\placefigure{qpofreq}

We first use our database of spectral parameters and corresponding 
0.5--10 Hz QPO parameters (for the definite QPOs) to explore the 
correlation between QPO frequency and 
flux from the disk black body over a larger range of light curves
than were studied by Markwardt, Swank, \& Taam (1999)
and Trudolyubov, Churazov, \& Gilfanov (1999). The results 
are shown in Figure \ref{qpofreq}.  Above 5 Hz, the QPO 
frequency increases slightly with disk flux, and at frequencies between 
0.5--5 Hz 
the QPOs generally occur at a nearly constant disk flux between 
0.2--0.5$\times 10^{-8}$ ergs/cm$^2$s. However, four of the variable 
observations (1996 October 7, 1997 August 14, 1997 September 9, and 1997
October 30) seem to lie off this track at frequencies below 4 Hz. These
exceptional data points are from dips during observations which exhibit 
infrared and radio flares (yellow-green points; see Figure \ref{lcurves}e) 
and observations when spectral analysis indicate the inner disk
has been evacuated (yellow-green points; see Figure \ref{lcurves}c).  
The QPO frequency is correlated with the power law flux during some
individual observations, even at lower fluxes. However, each 
observation traces its own track in Figure \ref{qpofreq}b, and it is difficult
to make a generalization of the relationship between power law flux
and QPO frequency that is valid for the entire set of observations. 

\placefigure{freqrt}

To further investigate the relationship between the 0.5--10 Hz QPO and the 
disk flux (which is simply proportional to $R_{col}^2 T_{col}^4$) we 
plot the QPO frequency versus the temperature and radius of the
inner disk in Figures \ref{freqrt}a and b respectively. It is apparent that 
the QPO frequency generally increases with the disk temperature, and 
decreases with disk radius. 
The observations with both high inner disk 
temperature and small radii (green squares and green asterisks) are 
exceptions to this trend; these points are off the 
temperature scale with $T_{col} > 3$ keV in Figure \ref{freqrt}a, and  
approach $R_{col} = 0$ in Figure \ref{freqrt}b.
We have compared the relationship between QPO frequency and photon index as
well (not shown), finding no apparent correlation between the 
two parameters.
For the majority of the observations (excepting the green squares and asterisks
with high $T_{col}$), the inner disk temperature is 
clearly the parameter that is most useful in predicting QPO frequency.

We next consider the width and the integrated amplitude of the 0.5--10 Hz
QPO as they relate to the spectral parameters.
The integrated fractional amplitude is
$$ A_{QPO} = \sqrt{{\pi W H} \over {2I}},$$
where $W$ is the width of the Lorentzian, 
$H$ is the maximum value of the Lorentzian in Leahy normalized units,
and $I$ is the mean count rate. 
The coherence parameter, $Q$ is a measure of the relative width of the QPO: 
$$Q = \nu/W,$$
where $\nu$ is the centroid frequency of the QPO.

\placefigure{qpoamp}
\placefigure{amprt}

Figure \ref{qpoamp}a shows that there is significant scatter in the 
correlations between QPO amplitude and disk flux, but that the QPOs
with the highest amplitude occur at low values of disk flux.
There is no significant correlation between QPO amplitude and power law 
flux in Figure \ref{qpoamp}b.
Figure \ref{amprt}a demonstrates that the amplitude of the QPO increases
with decreasing $T_{col}$. However, there is no such correlation between 
amplitude and $R_{col}$ (Figure \ref{amprt}b).

\placefigure{qpoq}

Finally, the coherence parameter of the QPO does not appear to be 
correlated with either component of the flux, and its average value is $Q 
\sim$10 (Figure \ref{qpoq}). Similarly, there is no correlation between either
$T_{col}$ or $R_{col}$ (not shown), in direct contrast to the
variations in the frequency and amplitude of the QPO.
We note however that features in the power 
spectrum that could be interpreted as very broad QPOs ($Q < 3$) have been
excluded from consideration here.

\subsection{Comparison of Spectra with and without QPOs}

\placefigure{bbpl}
We next turn our attention to the systematic changes that may occur when 
the 0.5--10 Hz QPO appears on or off. 
Figure \ref{bbpl} demonstrates that the correlation between disk black body 
flux and power law flux are strikingly
different when the QPO is present as opposed to when it is not.
When the QPO is present, the power law flux is much more variable than the 
disk flux. For the most part, the changes in the total flux track vertically
in this diagram with a relatively constant disk black body flux 
(0.3--1.3 $\times 10^{-8}$ ergs/cm$^2$s) and varying power
law flux (0.5--13 $\times 10^{-8}$ ergs/cm$^2$s). 
The points at the minimum power law flux correspond to
the lowest count rates during hard dips, such as those in 
Figure \ref{lcurves}c (yellow-green points), 
The vertical tracks are traced by the changing power law flux during the
entry into and exit from hard dips, such as those in Figures \ref{lcurves}e and
j (blue and purple points).
There are also several horizontal branches in Figure 
\ref{bbpl}a, which represent the change in disk flux that
occurs during the bright, hard emission such as in Figure \ref{lcurves}f 
(orange points). On the other hand, no steady observation 
(green and blue squares; Figure \ref{lcurves}a) that contains a QPO is seen
with a disk black body flux greater than $0.5 \times 10^{-8}$ ergs/cm$^2$s.
This indicates that when the 0.5--10 Hz QPO is present, changes in 
luminosity are basically confined to changes in the power law component.
Substantial increases in the disk flux only occur when the disk structure 
is cycling through unstable configurations. 

When the QPO is absent, the black body component is much more variable
than the power law.
In most cases, the flux in the power law component remains between 
2--5 $\times 10^{-8}$
ergs/cm$^2$s (less than half of the maximum), while the flux from the disk 
is seen to vary between 0.7--6 $\times 10^{-8}$ ergs/cm$^2$s.
The horizontal track in Figure 
\ref{bbpl}b corresponds to the soft emission that follows the hard dips 
during the observations similar to those in Figures \ref{lcurves}c and e
(the small yellow-green and blue points respectively). 
However, the soft dips of 1997 September 16 
(orange points; Figure \ref{lcurves}f), the variable high luminosity soft 
states (red points; Figures \ref{lcurves}g and h), 
and the soft-steady observations (red squares; Figure \ref{lcurves}b) 
also lie on this same track. Our interpretation of Figure \ref{bbpl}b
is that the absence of the 0.5--10 Hz QPO corresponds to an accretion mode
in which changes in the luminosity are primarily seen in the thermal
component of the spectrum.

However, one may wonder whether we do not detect a 0.5--10 Hz QPO in the
accretion mode when flux from the disk is most variable simply because of
the decrease in QPO amplitude with disk black body flux suggested by 
Figure \ref{qpoamp} (see also \cite{tcg99}). This is not an issue
at disk black body fluxes less than $2\times10^{-8}$ ergs s$^{-1}$ cm$^{-2}$,
since the QPOs which we find there have much higher amplitudes that
the upper limits for the ``no QPO'' group. At higher fluxes however, our 
upper limits ($\sim$ 0.6\%) are only a factor of a 
few smaller than the faintest QPOs which we can detect with certainty. 
Nonetheless, several indications lead us to believe that the 0.5--10 Hz 
QPO truly is absent along the whole horizontal branch in Figure \ref{bbpl}b.
We have visually inspected both dynamic PDS (as in Figure \ref{lcurves}) 
and individual PDS integrated for longer times (as in Morgan, Remillard, 
\& Greiner 1997) for all of our observations, and it is clear that the 
power spectrum differs dramatically during the two accretion modes. 
Furthermore, if we plot all of our data regardless of QPO properties 
(not shown), the two accretion modes which we report are still evident--- 
searching
for the 0.5--10 Hz QPO only serves to choose one branch or the other other. 
Since the branch without the QPO is continuous from low disk black body fluxes
(where the QPO can clearly be detected) to high fluxes, we feel it is 
reasonable to believe that the 0.5--10 Hz QPO is genuinely absent from 
the accretion mode in which we do not detect this QPO.

\placefigure{rt}

Having found a fundamental difference in the spectrum of GRS 1915+105
when the 0.5--10 Hz QPO is present and absent, we now examine how these 
changes manifest themselves in the soft X-ray component of the
spectrum. Figure \ref{rt} demonstrates
that the inner color radius and temperature of the disk are well
correlated, even when comparing wildly different observations. The correlation
is even more compelling if one ignores those points from segments with small
inner disk radii and large temperatures, many of which lie off beyond the 
extent of the $x$-axis at temperatures greater than 3 keV (green asterisks 
and green diamonds; compare Table \ref{obs}).
Plotted on
top of the data in Figure \ref{rt} is a line corresponding to the function 
$R_{col} = 39 T_{col}^{-2} + 22$, which represents a least-chi squares
fit of the dependence of radius on temperature (the fit excludes the 
observations with abnormally small $R_{col}$ and large $T_{col}$). 
We believe that the form of this relationship is more instructive than
the parameter values themselves, as these parameters may be systematically 
dependent on the methods used to model the energy spectrum (e.g. the 
choice of an absorption column) at low $T_{col}$, and because some of the
spectral evolution may be due to extrinsic changes, such as modifications
in the spectrum due to electron scattering. The 
$T_{col}^{-2}$ dependence of $R_{col}$ indicates a constant disk flux at 
low color temperatures, which is seen clearly in Figure \ref{bbpl}a.
Notice also that the color radius is observed to be 
relatively constant when the QPO is absent (Figure \ref{bbpl}b), 
although even then there is significant scatter in the data around our 
empirical correlation. 

\placefigure{plt}
\placefigure{alphat}

Finally, we examine the relationship between the inner 
disk and the power law component of the spectrum. Figure \ref{plt} 
demonstrates that when the 0.5--10 Hz QPO is present, the flux from the power 
law component generally increases from $0.5-15 \times 10^{-8}$ ergs/cm$^2$s 
as the color temperature of the inner disk increases from 0.7--1.9 keV. 
There are, however, exceptions to this trend.
The horizontal branch at $\sim 3 \times 10^{-8}$ ergs/cm$^2$s of power law
flux is composed of the ringing
observation on 97 May 26 (Figure \ref{lcurves}d) and the ringing portion of
the observation on 97 May 18 (Figure \ref{lcurves}j). There is a second 
horizontal branch at $\sim 9 \times 10^{-8}$ ergs/cm$^2$s of power law 
flux which is corresponds to the observations with unusually 
small inner disk radii and large temperatures (green squares and green 
asterisks), and extends the scale of the plot at high temperatures. 
When the QPO is absent, the power law flux is generally weaker, and 
there is little correlation
between the power law flux and color temperature.

Figure \ref{alphat} investigates the relationship between power law index
and disk color temperature. When the QPO is present
the photon index of the 
power law component increases from 2.1--3.0 as the color temperature of 
the inner disk increases from 0.7--2.0 keV. 
The observations with unusually small inner disk radii lie off the plot
at $T_{col} > 3.0$ keV and $2.6< \Gamma < 3.1$.
When the QPO is absent,
there is no correlation between the temperature of the inner disk and the 
photon index. 

There are related 
correlations between the inner radius of the disk and the flux and 
photon index of the power law,
as would be expected given the correlation between radius and temperature,
but these correlations (not shown) are weaker. We therefore
conclude that when the 0.5--10 Hz QPO is present, the temperature of the 
inner disk can be used to roughly predict the photon index and the flux from 
the power law. 

\section{Discussion}

Several of the results of this paper demonstrate the significance of the 
0.5--10 Hz QPO in GRS 1915+105. 
First, we have shown that the frequency of the QPO is best 
correlated with the temperature of the inner
accretion disk (Figure \ref{freqrt}). This indicates that the time scale of 
this QPO is set by conditions in the optically thick accretion disk.

Second, we have discovered that when the 0.5--10 Hz QPO is present,
changes in the 
luminosity of GRS 1915+105 occur mainly in the power law component,
and that when the QPO is absent changes in luminosity occur in 
the thermal emission from the inner accretion disk (Figure \ref{bbpl}). 
Chen, Swank, \& Taam (1997) and Markwardt, Swank, \& Taam (1998) have similarly
noted that the 0.5--10 Hz QPO is characteristic of hard spectral states. 
Moreover, when the QPO is absent during the steady observations 
(where good statistics are available) we find that the hard X-ray component 
must be modeled with an exponential that produces
negligible flux above $\sim$50 keV, suggesting that the mechanism generating 
the Comptonizing electrons is inhibited or that the electrons have
been quenched by Compton cooling. 
These results further establish the link between the 0.5--10 Hz
QPO and the hard X-ray power law component in GRS 1915+105, as Morgan, 
Remillard, \& Greiner (1997) \markcite{mrg97} 
have already demonstrated that the fractional amplitudes of four QPOs 
increase with photon energy, extending well past the 
thermal component of the spectrum.

Two general classes of models for the formation of QPOs are relevant in 
seeking an understanding of the intimate relationship between the 0.5--10 Hz 
QPO and both
components of the X-ray spectrum of GRS 1915+105. First, numerous 
authors have proposed that oscillations in the inner accretion disk
could lead to QPOs (e.g. \cite{ct94}, \cite{act95}), 
and it has recently been noted that if these oscillations 
generate ``seed'' photons for Comptonization, 
the QPO may be predominately exhibited in the hard portion 
of the energy spectrum (\cite{st98}). However, the 0.5--10 Hz QPO in 
GRS 1915+105 disappears when the disk appears stable (in terms of the color 
radius, as in Figure \ref{rt}b) and luminous (Figure \ref{bbpl}), and there 
is no clear reason why the hypothesized
disk oscillations should cease at these times.

The second class of models postulates that QPOs may form at 
a geometric boundary between the
optically thick disk and the population of Comptonizing electrons, allowing
for simultaneous modulations of both the hard and soft components of
the spectrum. Models which involve a shock front 
between the optically thick and thin regions of the accretion 
flow can provide such effects. Oscillations in the height and width of the 
shock could result in QPOs by modulating either the number
of ``seed'' photons or the populations of Comptonizing electrons 
(\cite{kht97}). However, the time scales of such oscillations are on the order
of $\sim$100 Hz, and better serve as a explanation of the 67 Hz feature seen 
in GRS 1915+105 (\cite{tlm98}). Nonetheless, numerical simulations by 
Molteni, Sponholz, \& Chakrabarti (1996) \markcite{msc96}
indicate that oscillations in the radial position of 
a shock could generate a $\sim$1--10 Hz QPO. These oscillations 
are based on a resonance between the cooling time of the shock and
in the free fall time of matter into the black hole, so the time scale is set 
by the radius at which the shock forms. The radius is
in turn is set by the accretion rate in the disk, so the correlation between
$T_{col}$ ($\propto ({\dot M}/R^3)^{1/4}$) and QPO frequency in Figure 
\ref{freqrt} may support some aspects of this model. 

In order to most simply explain the fact that the power law is only 
active when the 0.5--10 Hz QPO is present, 
one may hypothesize that the same mechanism 
generates both the QPO and the relativistic electrons responsible for the hard 
X-rays in GRS 1915+105. On the other hand, a 9--30 Hz QPO with similar 
spectral characteristics in GRO J1655-40 {\it decreases} 
in frequency with X-ray luminosity and disk temperature
(\cite{rem98}, \cite{sob99a}), precisely the 
opposite of the QPO is GRS 1915+105 (Figures \ref{qpofreq} and \ref{freqrt}). 
If some
QPOs are indeed part of the mechanism generating Comptonizing electrons, 
it is not clear why such a discrepancy would exist between two apparently
similar sources.

The results contained in Figure \ref{rt} contribute to further reasons for
caution in using the disk color radius to 
estimate the radius of the last stable orbit in accreting black hole
systems (\cite{zcc97}).
We have demonstrated that the disk color radius approached a stable value
only when the 0.5--10 Hz QPO is absent. If we omit the points at $T_{col} 
< 1.2$ keV as likely statistical ``leaks'' which are mis-categorized as 
lacking a QPO, we 
find a minimum stable color radius, $R_{col,min} = 44 \pm 8 (1\sigma)$ 
km (the uncertainty here is standard deviation of the values for $R_{col,min}$
used to compute the average,
which by coincidence is equal to our average uncertainty on a single 
measurement of $R_{col}$). Note that even after this careful selection 
of data, there is still a 20\% scatter in color radius values, which should be 
considered in addition to the uncertainty on the particular correction factors
from which a physical radius is estimated from the values of $R_{col}$.
If we now use Equation 3 in Zhang, Cui, \& Chen (1997)\markcite{zcc97} 
along with their correction factors to convert our estimate of 
$R_{col,min}= 44 \pm 8$ to an estimate of the 
last stable orbit, we find the compact object in GRS 1915+105 has a 
mass $M \sim 14 \pm 4 M_{\odot}$ km for a 
Schwarschild black hole, $M \sim 65 \pm 20 M_{\odot}$ for a prograde Kerr 
black hole, and $M \sim 9 \pm 3 M_{\odot}$ for a retrograde Kerr black hole.
However, we note that significant uncertainty in this estimate of the mass 
is introduced not only through the unknown spin period, but also through 
systematic diffuculties in determining the disk parameters (e.g. 
\cite{sob99a}).

The radio-loud hard-steady observations with $R_{col} < 10$ km deserve
further consideration, as they provide exceptions to 
many of the global trends in Figures \ref{qpofreq}--\ref{alphat}. 
In addition to the exceptional values of $R_{col}$ and $T_{col}$ in 
Table \ref{fits}, these observations exhibit
particularly strong iron emission in the energy spectrum ($N_{gauss} >
8.5\times 10^{-2}$). Moreover, these radio-loud hard-steady observations are
distinct from the radio-quiet observations, in that joint PCA/HEXTE fits reveal
no necessity for either a reflection component or a cut-off in the power law 
(Table \ref{hexte}). The time intervals when the radio-loud hard-steady 
observations occurred (1996 July--August and 1997 October) have also been
singled out as exhibiting optically thick radio emission, and the intervals 
are referred to as the ``plateau'' radio state by (\cite{fen99}; see also 
references therein). It is evident that these 
time intervals deserve additional attention, not only to explore the
limits of the the multi-temperature disk model in the X-ray band, but also 
to elucidate the relationship between emission from GRS 1915+105 in the X-ray 
and radio wavelengths.

Finally, our accretion modes may be compared with the ``states'' which Belloni 
et al. (1997b) via Figure \ref{bbpl} used to characterize the variable
emission from GRS 1915+105. The ``quiescent'' state of Belloni et al., which
corresponds to hard dips such as those in Figure \ref{lcurves}c, is
associated with the accretion mode that exhibits the 
0.5--10 Hz QPO (Figure \ref{bbpl}a; see also the discussion in Section
3.2). 
The ``flaring'' state of Belloni et al. describes the soft, bright emission 
that follows the hard dips (Figure \ref{bbpl}a), and is associated with 
the accretion mode without the QPO. We also find that the fluxes taken 
from steady observations (large squares in \ref{bbpl})
when the emission from GRS 1915+105 was stable for entire orbits 
($> 3000$ s) lie on very similar tracks as the fluxes derived from 32 s 
intervals (small points) when the source was highly variable.
This suggests an alternative perspective for understanding the cyclic
variations in GRS 1915+105. The variations may represent transitions 
between two quasi-steady accretion modes, signified by the presence or 
absence of the 0.5--10 Hz QPO. This concept may be useful
toward understanding the ``reverse'' dip cycles represented in Figure
\ref{lcurves}f (orange points in Figure \ref{bbpl}), which are not understood
in the context of the CV-like disk instability model (\cite{bel97a}).
The physics underlying these accretion modes is
unknown, but the transitions may require an explanation beyond the traditional
thermal disk instability.

\section{Conclusions}

We have analyzed a set of 27 PCA observations of GRS 1915+105 which are
a representative sample of the spectral shapes and variability patterns
from this source. We modeled the energy spectrum with a disk black
body, a Gaussian, a constant interstellar absorption, and either a power
law or exponential spectrum. We also searched the PDS for a 0.5--10
Hz QPO. Finally, we compared the parameters of this QPO with the spectral 
parameters, and separated our database into two groups based upon whether 
or not they contained this QPO.

We extended the results of Markwardt, Swank, \& Taam (1999) and 
Trudolyubov, Churazov, \& Gilfanov (1999)
by demonstrating that the 0.5--10 Hz
QPO frequency increases as the color temperature of the inner disk 
increases from 0.7 keV--1.5 keV and as the color radius 
decreases from 120--20 km (Figure \ref{freqrt}). 
On average, the fractional RMS amplitude of the QPO 
decreases from 25\% to less than 3\% as the temperature of the inner 
disk increases (Figure \ref{amprt}), while the coherence of the QPO does not 
change systematically with any of the spectral parameters (Figure \ref{qpoq}). 

We have also demonstrated that the 0.5--10 Hz
QPO serves as a marker for two distinct 
accretion modes (Figure \ref{bbpl}). When the QPO is present, 
accretion energy is channeled primarily into the 
power law component of the spectrum and the power law flux and spectral index 
roughly increase as the temperature of the disk increases. When the QPO is 
absent, accretion energy is primarily expressed in the disk black body 
component, while the power law flux and photon index are largely uncorrelated
with the disk color temperature (Figures \ref{bbpl}, \ref{plt}, and 
\ref{alphat}). 

The color radius and 
temperature of the inner accretion disk are related by 
$R_{col} \propto T_{col}^{-2} + const$, and that when the QPO is absent, the
color radius remains near a minimum value of $R_{col,min}= 44 \pm 8 
(1 \sigma)$ km. 
Assuming this minimum 
value of the color radius is indicative of the innermost stable orbit, 
rough estimates of 
the mass of the black hole in GRS 1915+105 range between $\sim 6-80$ 
M/M$_{\odot}$, depending on the spin.

Eleven of the 27 observations require particular attention in modeling their 
spectra. Six observations (with a hard spectrum and 0.5--10 Hz QPO) are
characterized by relatively high radio flux, high color temperatures, and 
color inner disk radii less than 5 km. The reduced chi-squared values for 
the spectral fits are good, but there is no physical interpretation of 
the dramatic decrease in the normalization of the thermal component. 
In five other observations that have soft spectra and lack the 
narrow 0.5--10 Hz QPO, the hard component must be modeled with 
an exponential function.

Finally, we find that much of the emission from GRS 1915+105 can be reduced to
two accretion modes, distinguished by the presence or absence of 
a 0.5-10 Hz QPO.

\acknowledgements

We thank Hale Bradt, Al Levine, Wei Cui, and Elizabeth Waltman for careful 
readings of drafts of this paper, and Craig Markwardt, Jean Swank, 
Ronald Taam for discussions about the content of this paper. This work 
was supported by NASA contract NAS 5-30612.

\begin{figure}
\figurenum{1}
\plotone{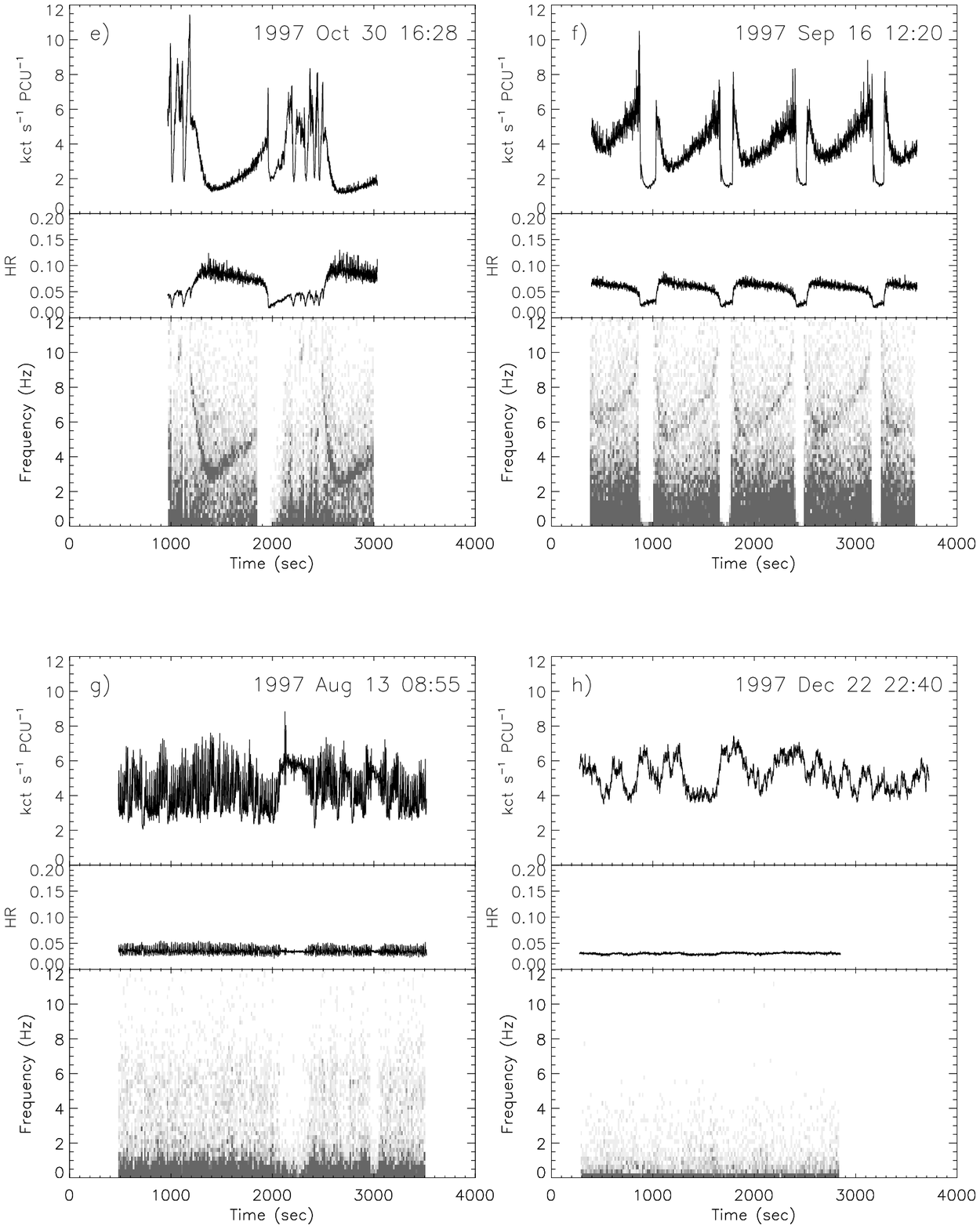}
\caption{Figure 1, continued}
\end{figure}

\begin{figure}
\figurenum{1}
\plotone{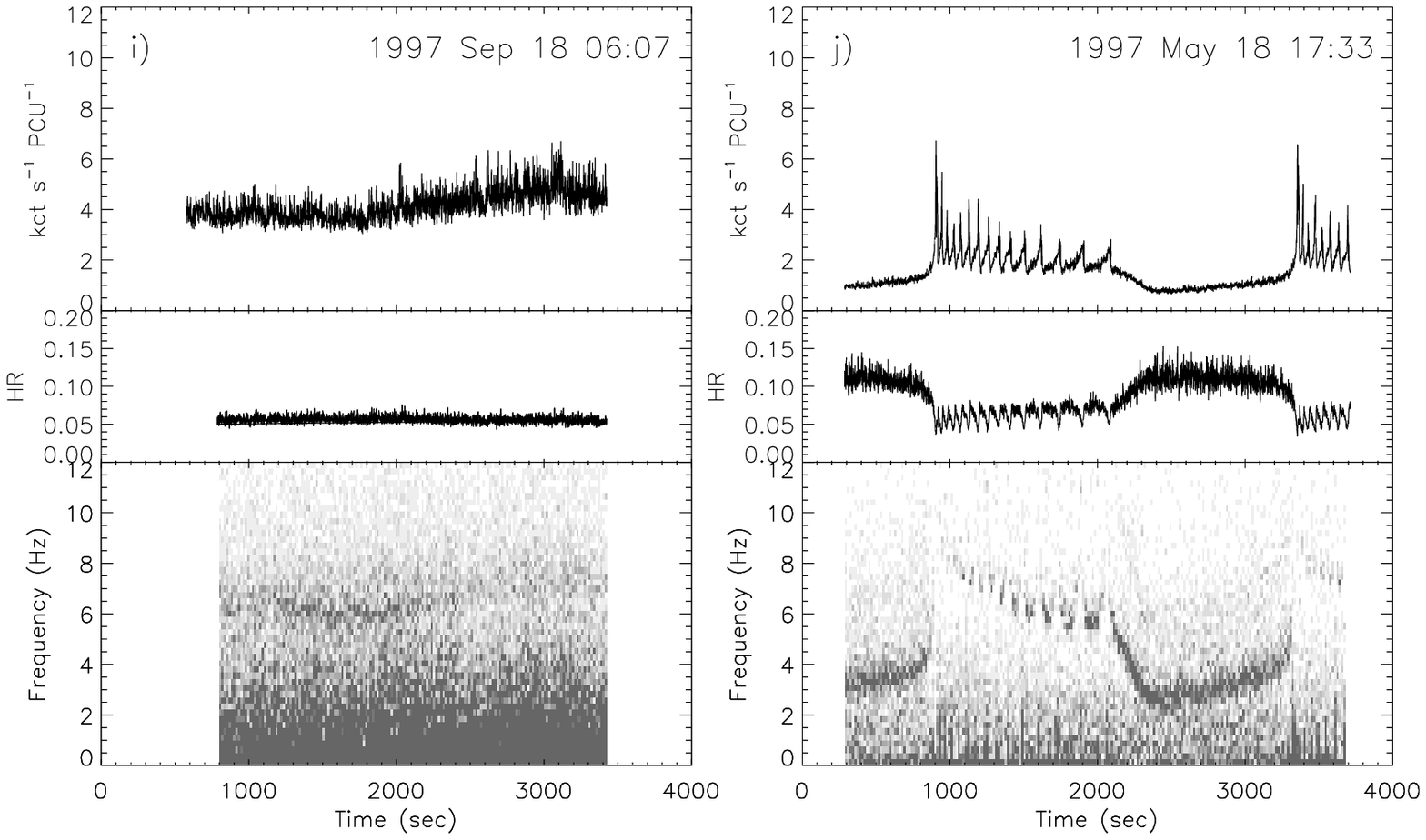}
\caption{Figure 1, continued}
\end{figure}

\begin{figure}
\plotone{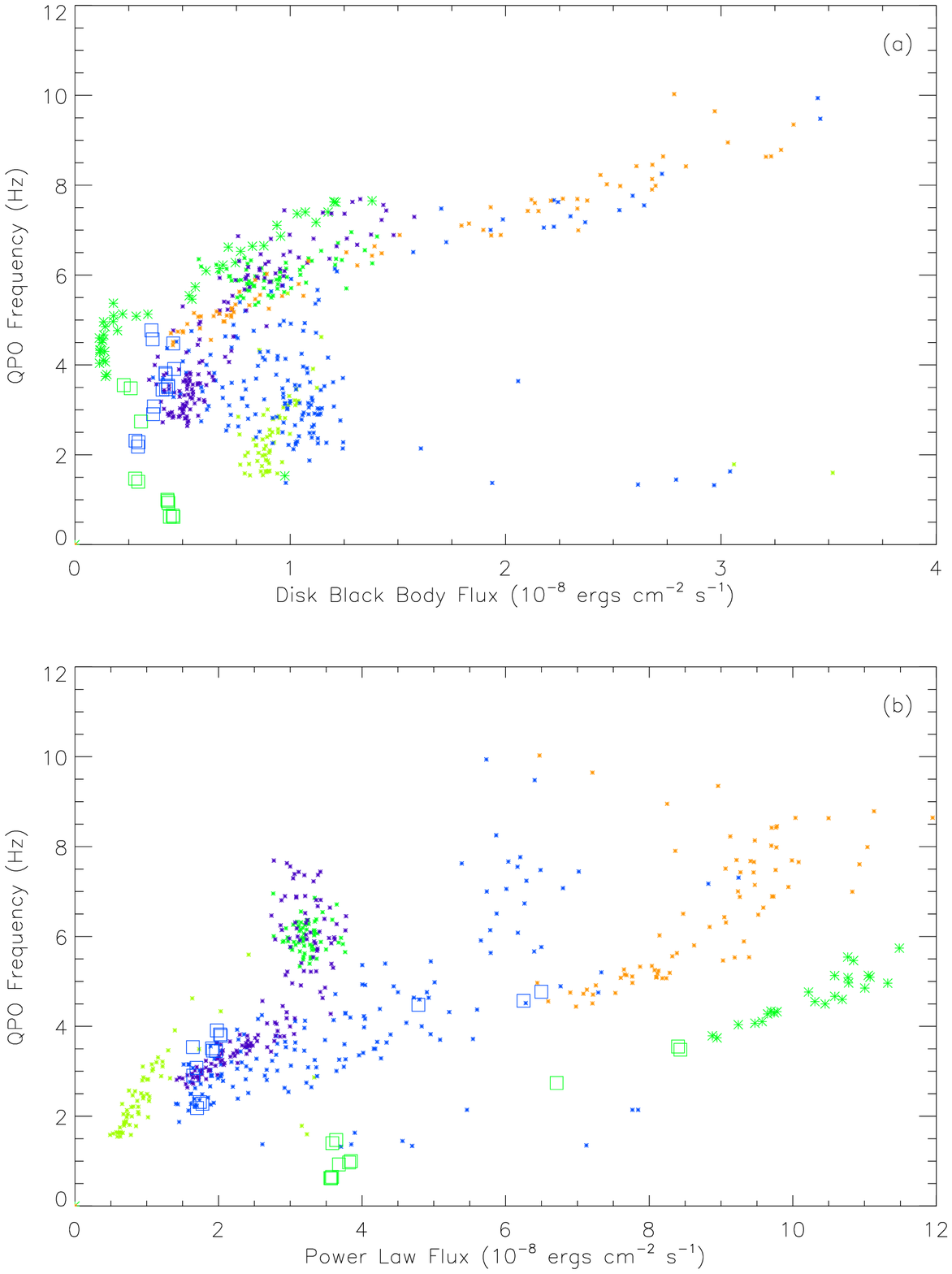}
\caption{QPO frequency vs. thermal flux from the disk (a) and vs.
flux from the power law (b). The key to the symbols is described in 
the text.}
\label{qpofreq}
\end{figure}

\begin{figure}
\plotone{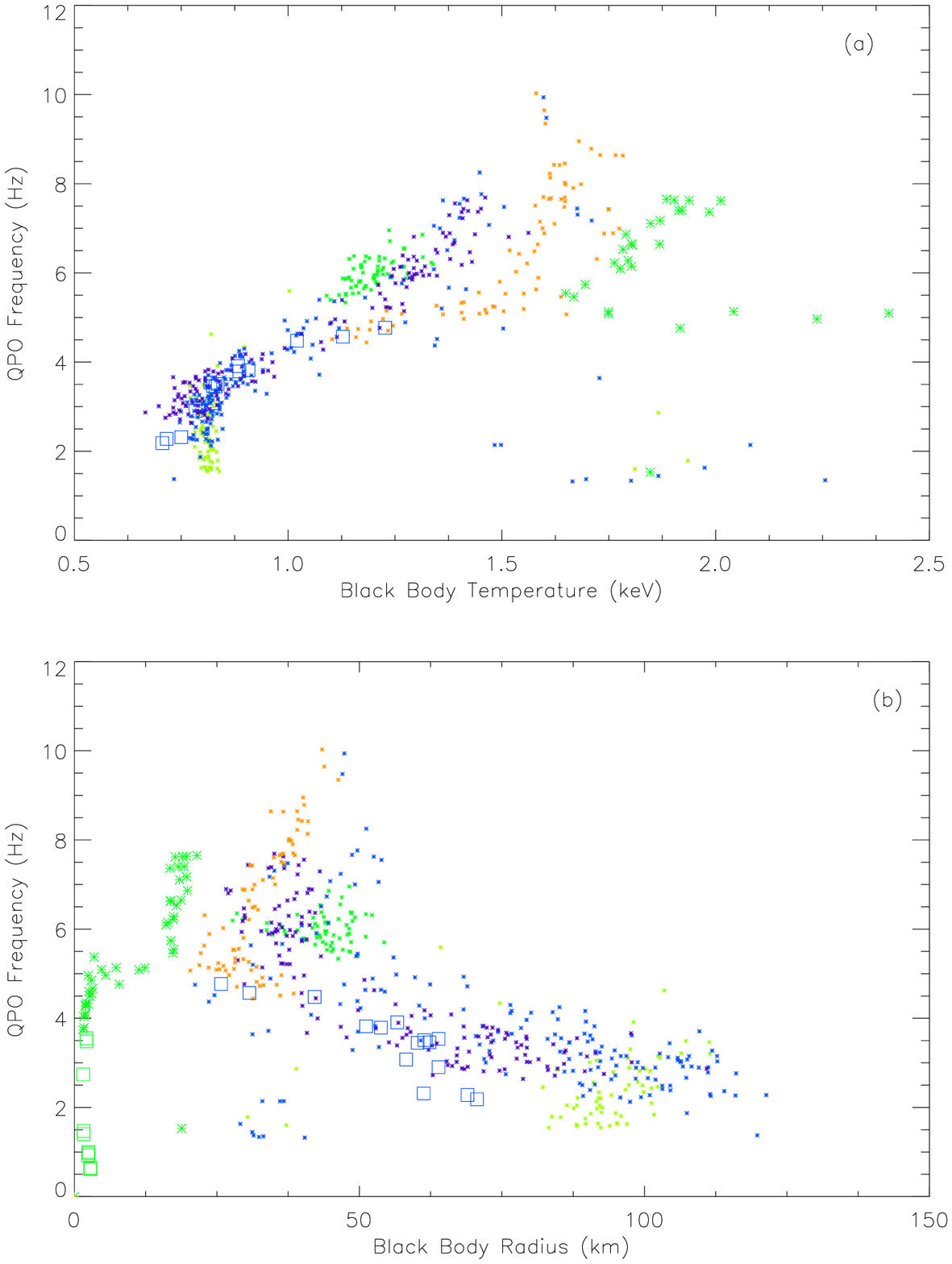}
\caption{QPO frequency vs. the color temperature of the inner
disk (a) and vs. the color radius of the inner disk (b). The points 
with abnormally high temperature (green squares and asterisks) are off 
the scale of panel a. The symbol key is described in the text.}
\label{freqrt}
\end{figure}

\begin{figure}
\plotone{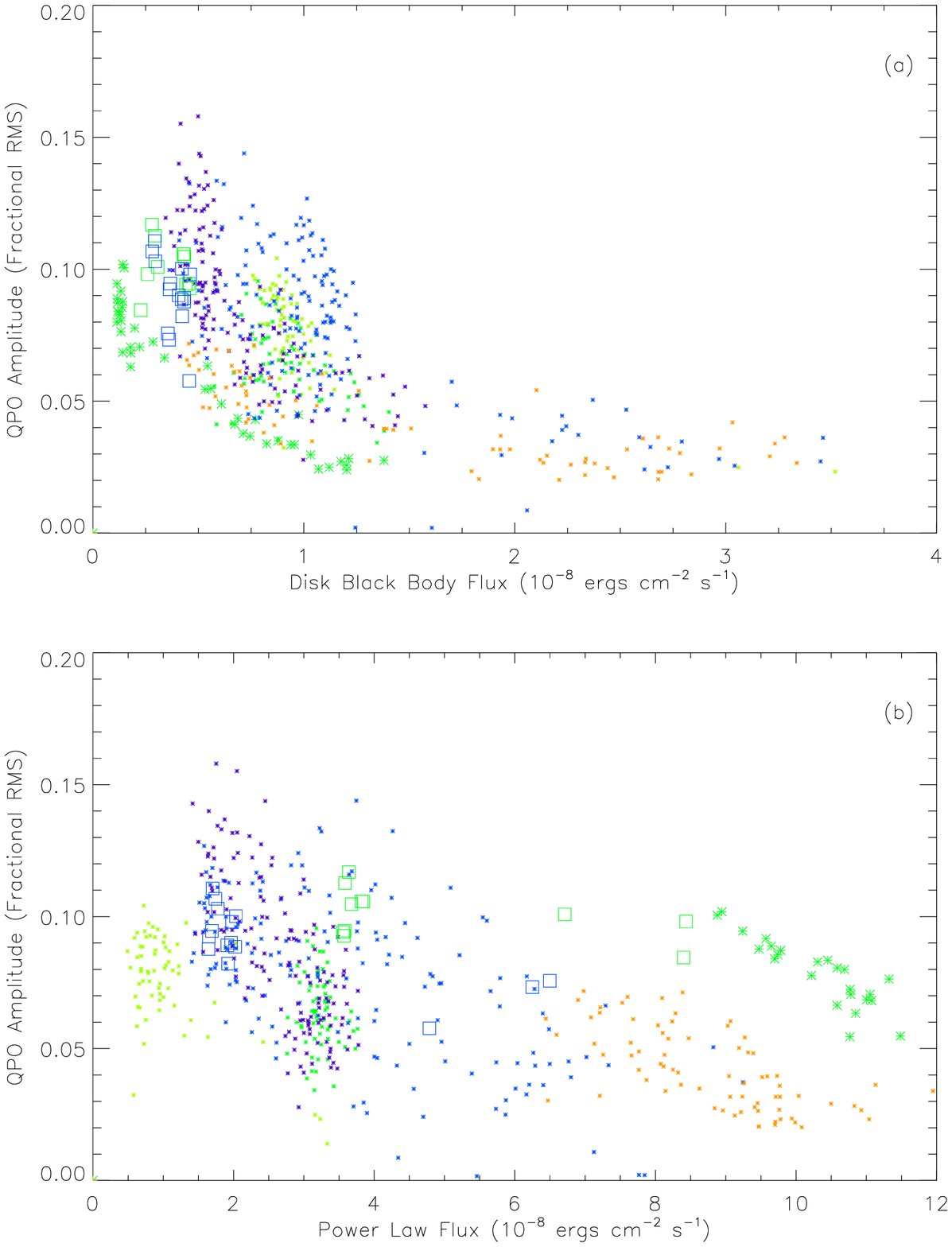}
\caption{RMS QPO fractional amplitude vs. thermal flux from the 
disk (a) and vs. flux from the power law (b). The symbol key is described
in the text.  }
\label{qpoamp}
\end{figure}

\begin{figure}
\plotone{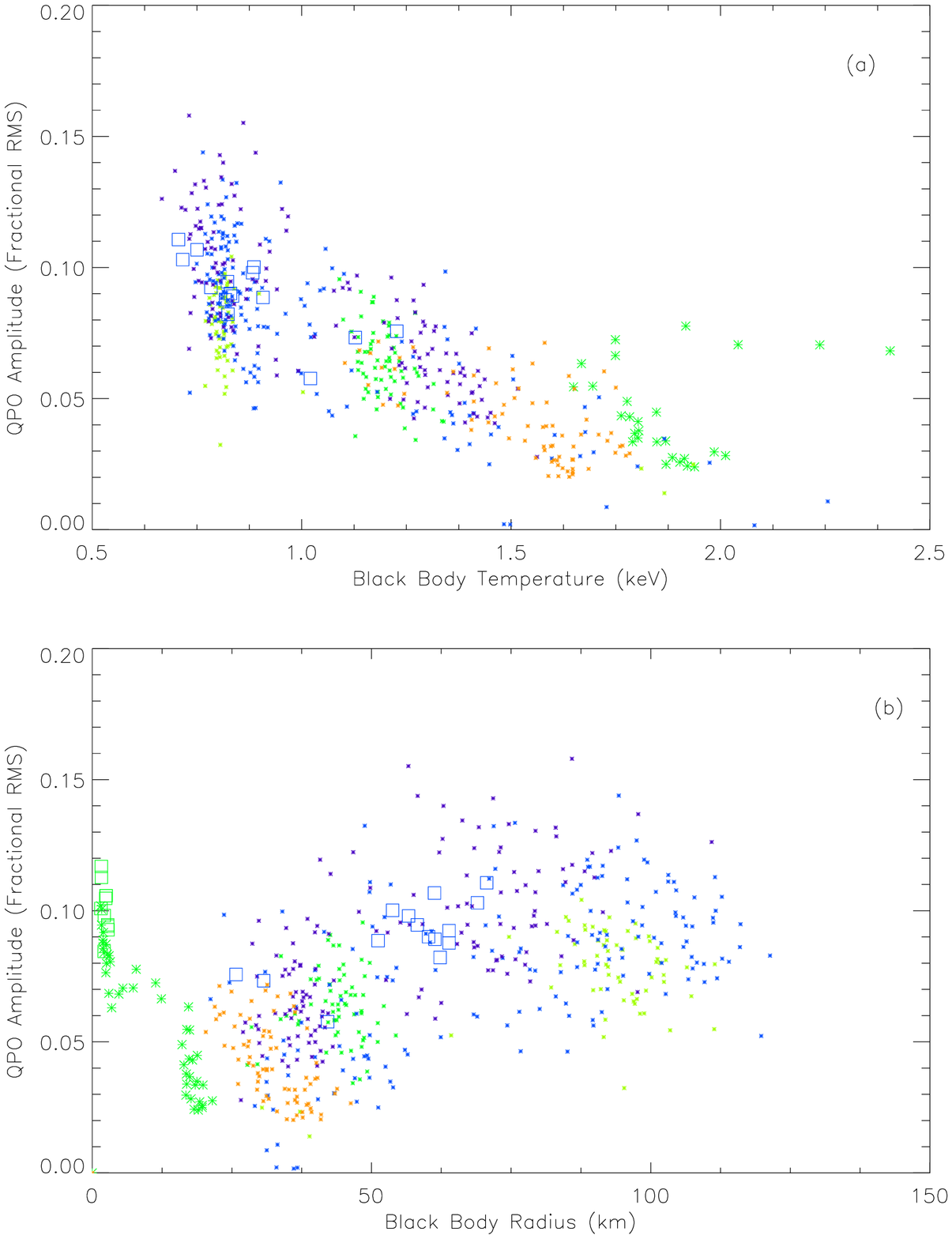}
\caption{QPO amplitude vs. the color temperature of the inner
disk (a) and vs. the color radius of the inner disk (b). The points with 
abnormally high temperature are off the scale of panel a. The symbol key is 
described in the text.}
\label{amprt}
\end{figure}

\begin{figure}
\plotone{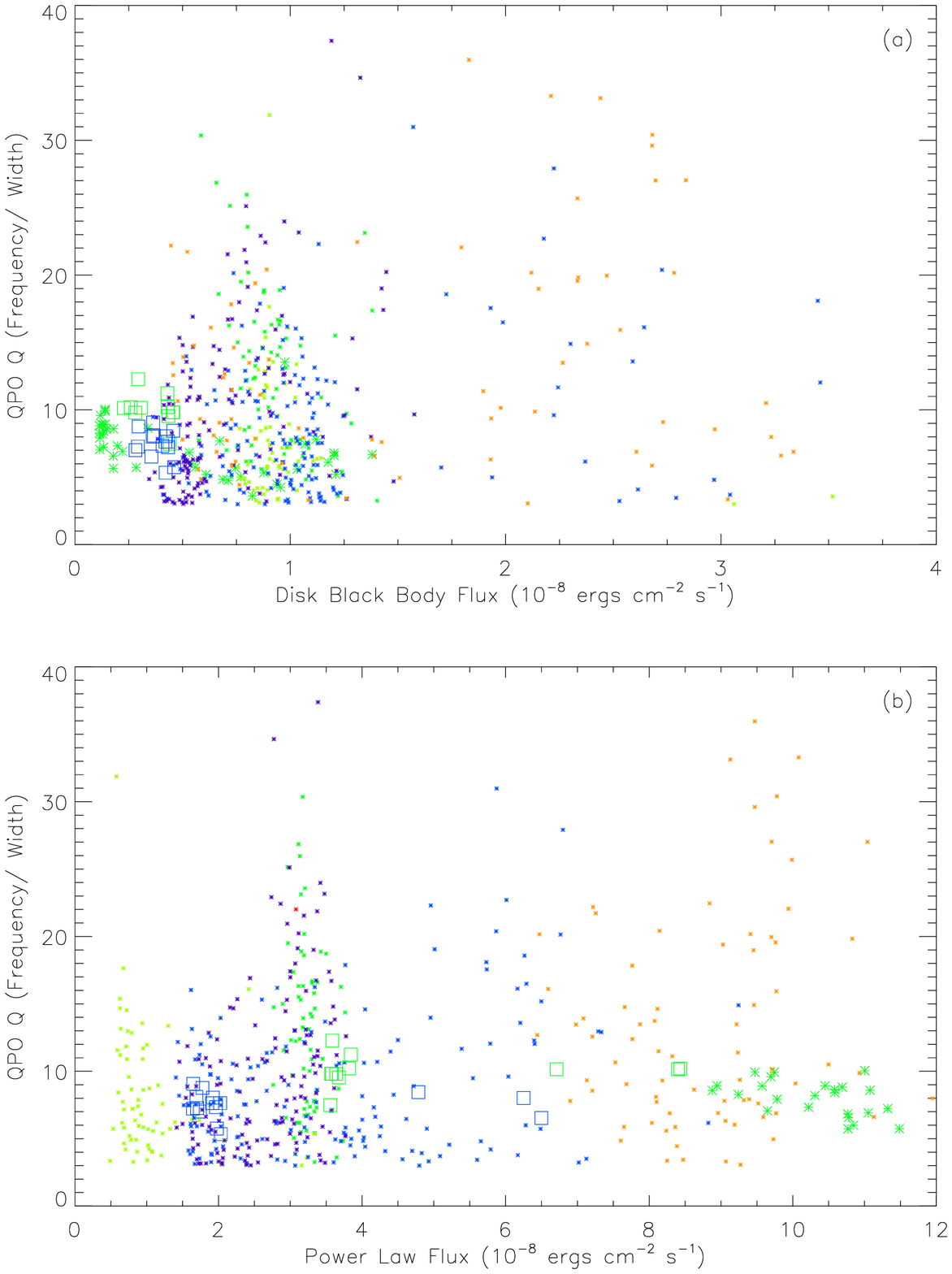}
\caption{QPO coherence ($Q=\nu/\delta \nu$) vs. thermal flux from the 
disk (a) and vs. flux from the power law (b). The symbol key is described 
in the text.}
\label{qpoq}
\end{figure}

\begin{figure}
\plotone{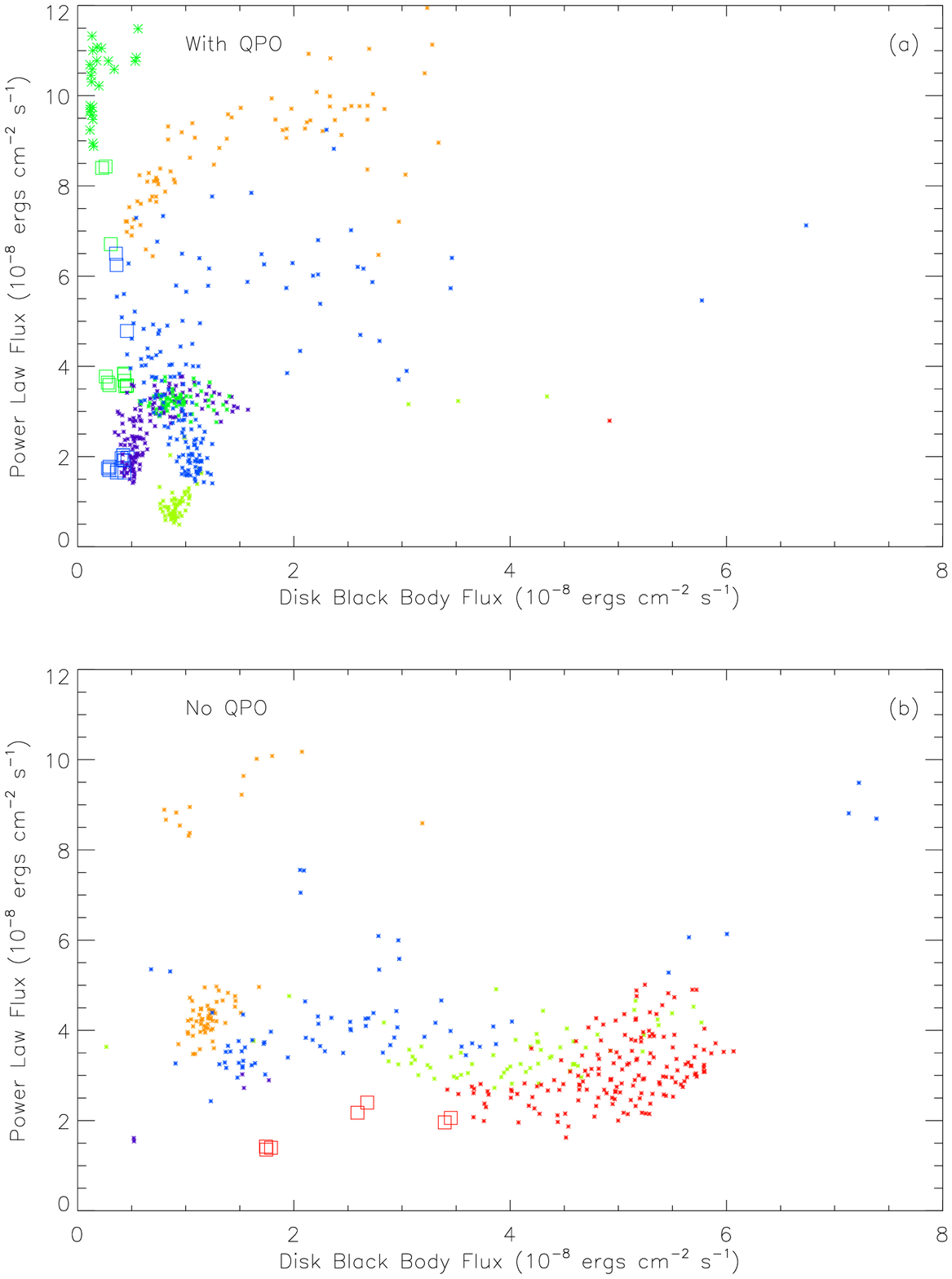}
\caption{Flux from the power law vs. the thermal flux from 
the disk when the 0.5-10 Hz QPO was present (a) and when the same QPO was
absent (b). The symbol key is described in the text.}
\label{bbpl}
\end{figure}

\begin{figure}
\plotone{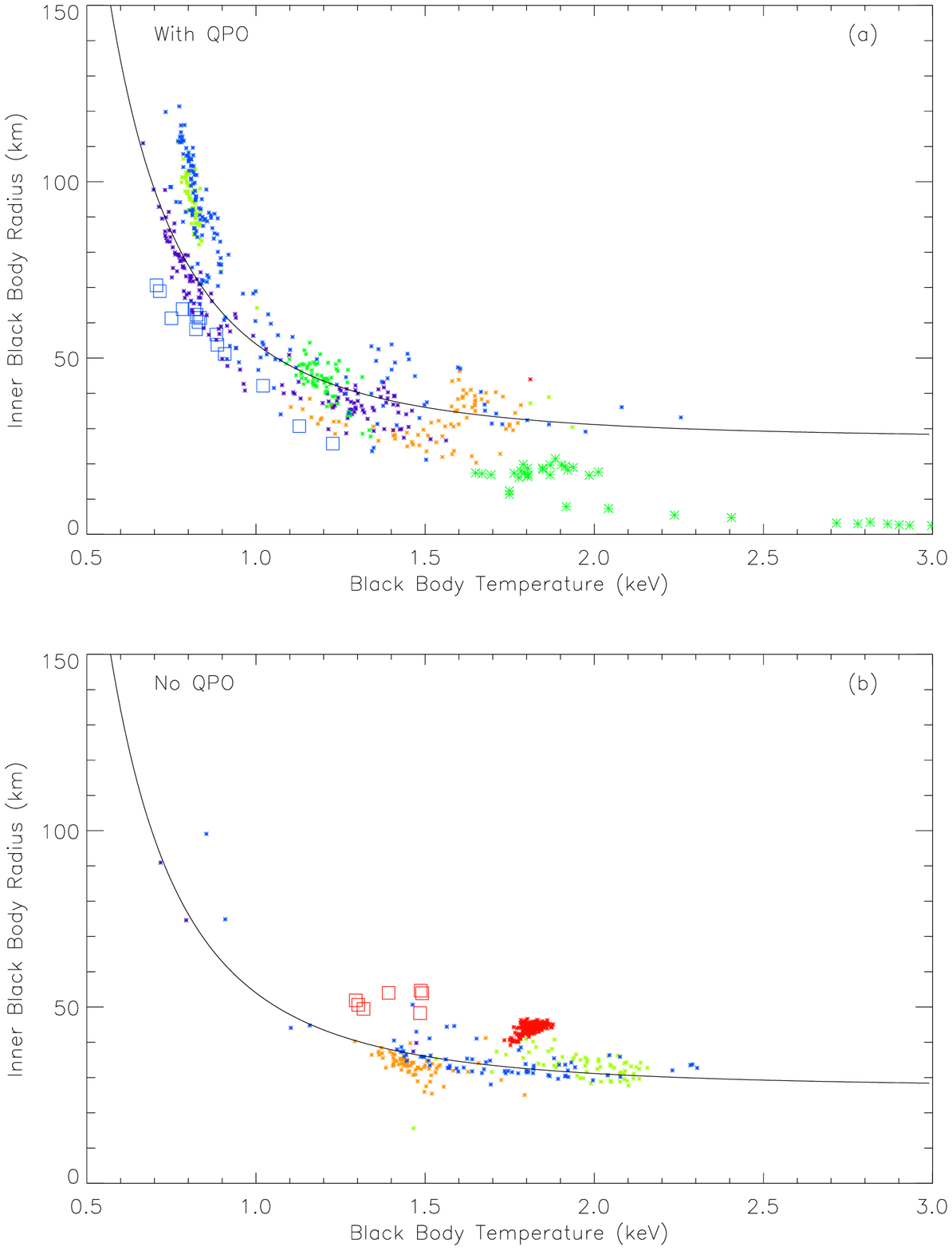}
\caption{Plots of the color radius of the inner disk vs. the color temperature 
of the inner disk when the 0.5-10 Hz QPO was present (a) and when
the same QPO was absent (b). The lines drawn 
correspond to $R_{col} = 39T_{col}^{-2} + 22$. The points with abnormally
high temperature are off the scale of panel a.
The symbol key is described in the text.}
\label{rt}
\end{figure}

\begin{figure}
\plotone{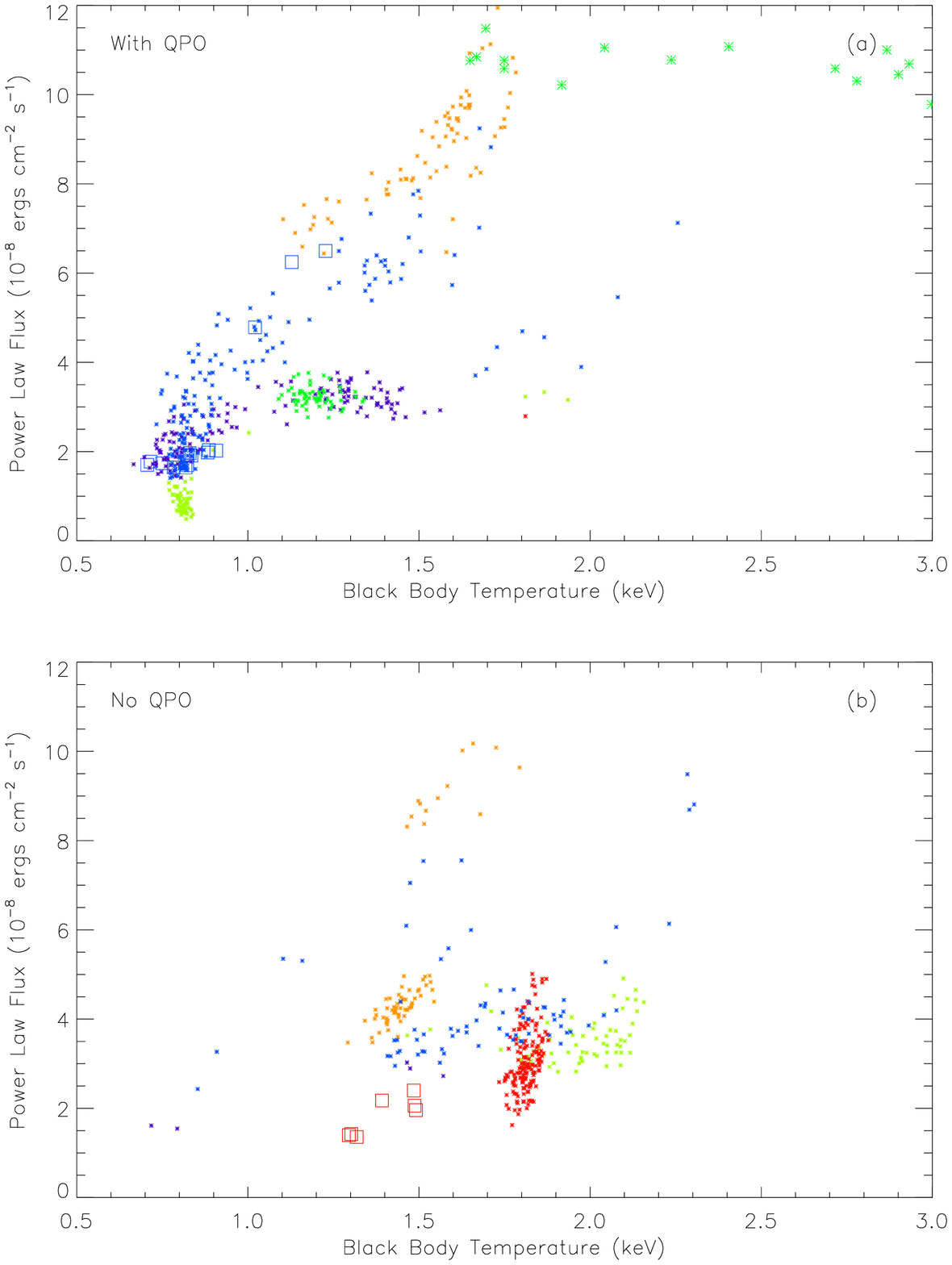}
\caption{Flux from the power law vs. the color temperature of the
inner disk when the 0.5-10 Hz QPO was present (a) and when the same QPO was
absent(b) . The points with abnormally high temperature are off the scale of
panel a. The symbol key is described in the text.}
\label{plt}
\end{figure}

\begin{figure}
\plotone{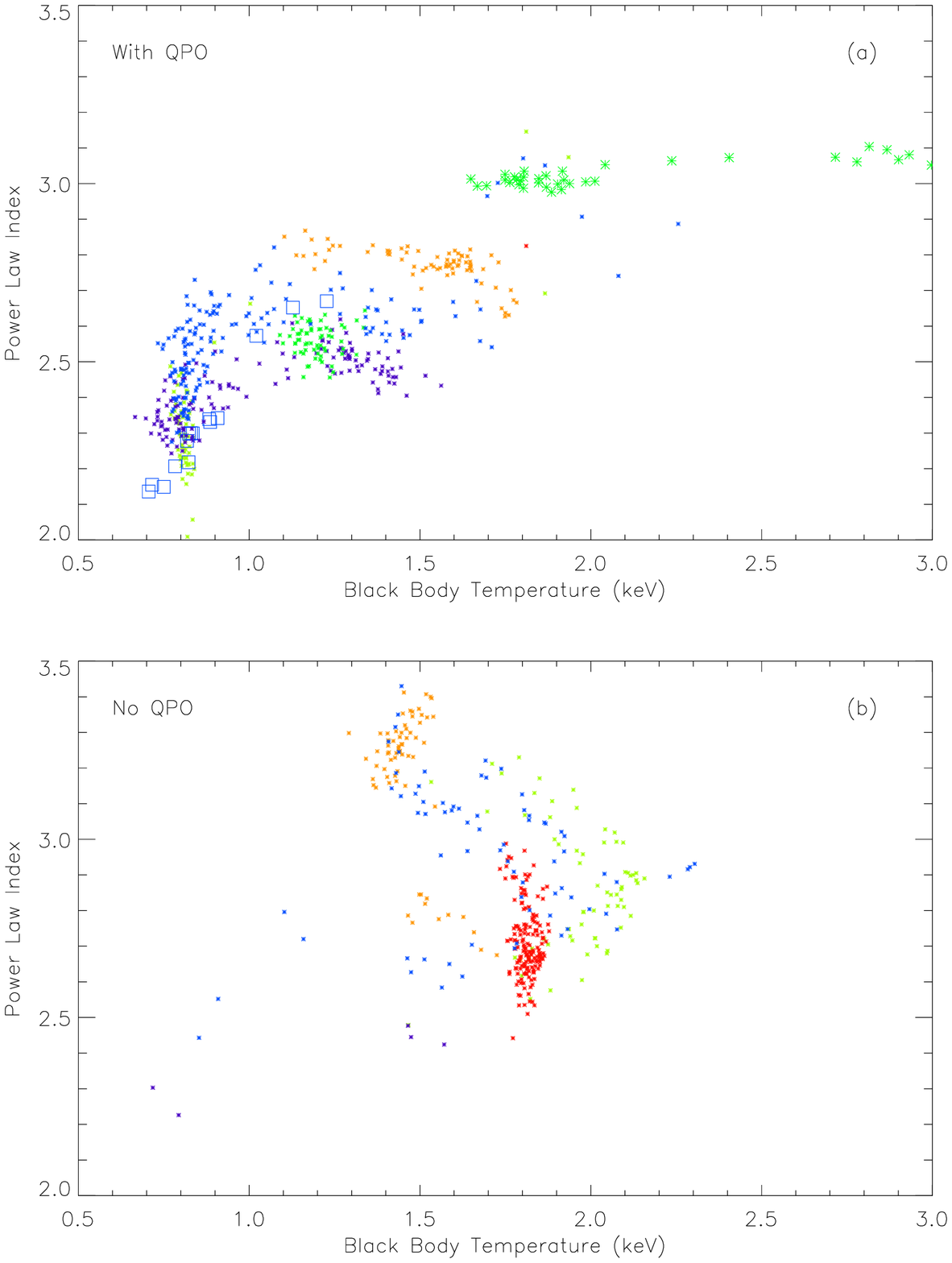}
\caption{Photon index vs. the color temperature of the inner
disk when the 0.5-10 Hz QPO was present (a) and when the same QPO was
absent (b). The points with abnormally high temperature are off the scale 
of panel a. The symbol key is described in the text.}
\label{alphat}
\end{figure}

\begin{deluxetable}{llrrcccc}
\scriptsize
\tablecolumns{8}
\tablewidth{0pc}
\tablecaption{Summary of Selected Observations \label{obs}}
\tablehead{
\colhead{Observation ID} 	& \colhead{Date} 	& \colhead{Exposure} & 
\colhead{C/s}	& \colhead{Variability}	& \colhead{Interval}
&\colhead{Radio Flux \tablenotemark{a}}		
&\colhead{QPO \tablenotemark{b}} \\
\colhead{} 			& \colhead{(UT)}	& \colhead{(sec)} &
\colhead{per PCU}	 	& \colhead{($\sigma/$mean} 	& \colhead{}
&\colhead{(Jy)} &\colhead{} 
}
\startdata
10408-01-06-00 & 1996 May 05 14:03 & 10020 & 3264 & 0.077 & orbit & \nodata & No\tablenotemark{d} \nl
10408-01-22-00	& 1996 Jul 11 02:16 & 2700 & 2107 & 0.038 & orbit & 0.050\tablenotemark{c} & Yes \nl
10408-01-27-00	& 1996 Jul 26 13:57 & 9420 & 1777 & 0.083 & orbit & 0.090\tablenotemark{c} & Yes  \nl
10408-01-28-00	& 1996 Aug 03 12:56 & 11280 & 1723 & 0.079 & orbit & 0.100\tablenotemark{c} & Yes \nl 
10408-01-38-00	& 1996 Oct 07 05:44 & 11280 & 3337 & 0.572 & 32 s & 0.003\tablenotemark{c} & Yes \nl
20402-01-04-00 & 1996 Nov 28 03:01 & 7020 & 2142 & 0.126 & orbit & \nodata & Yes \nl
20402-01-13-00	& 1997 Jan 29 20:49 & 10620 & 952 & 0.076 & orbit & 0.007 & Yes \nl 
20402-01-14-00	& 1997 Feb 01 21:01 & 9900 & 922 & 0.075 & orbit & 0.006 & Yes \nl
20402-01-15-00	& 1997 Feb 09 18:35 & 10500 & 831 & 0.086 & orbit & \nodata & Yes \nl
20402-01-16-00	& 1997 Feb 22 21:08 & 6060 & 821 & 0.077 & orbit & 0.014 & Yes \nl
20402-01-21-00 & 1997 Mar 26 20:00 & 3660 & 805 & 0.073 & orbit & 0.003 & Yes \nl
20402-01-28-00 & 1997 May 18 16:21 & 8100 & 1521 & 0.451 & 32 s & \nodata & Yes \nl
20402-01-30-00 & 1997 May 26 11:55 & 4260 & 1836 & 0.298 & 32 s & 0.010 & Yes \nl
20402-01-38-00 & 1997 Jul 20 10:23 & 7620 & 3853 & 0.113 & orbit & 0.009 & No\tablenotemark{d} \nl	
20186-03-03-00 & 1997 Aug 13 04:07 & 9900 & 4540 & 0.270 & 32 s & 0.013 & No\tablenotemark{d} \nl
20186-03-03-01	& 1997 Aug 14 04:08 & 9900 & 3567 & 0.654 & 32 s & 0.050 & Yes \nl
20402-01-41-00 & 1997 Aug 19 05:50 & 2340 & 4880 & 0.159 & 32 s & 0.009 & No \nl
20402-01-45-03	& 1997 Sep 09 06:01 & 10800 & 2539 & 0.527 & 32 s & 0.051 & Yes \nl
20186-03-02-03	& 1997 Sep 16 08:17 & 20280 & 3606 & 0.365 & 32 s & 0.105 & Yes \nl
20186-03-02-06 & 1997 Sep 18 03:00 & 27360 & 3504 & 0.300 & 512 s & 0.019 & Yes \nl
20402-01-49-01	& 1997 Oct 09 09:28 & 4900 & 1796 & 0.043 & orbit & \nodata & Yes \nl
20402-01-52-00	& 1997 Oct 25 06:27 & 18180 & 1543 & 0.057 & orbit & 0.051 & Yes \nl
20402-01-52-01 & 1997 Oct 30 16:27 & 2100 & 3130 & 0.613 & 32 s & 0.156 & Yes \nl
20402-01-55-00 & 1997 Nov 17 05:25 & 8800 & 3554 & 0.080 & orbit & 0.028 & No\tablenotemark{d} \nl
20402-01-60-00 & 1997 Dec 22 21:40 & 12540 & 5436 & 0.159 & 32 s & 0.013 & No \nl
30402-01-04-00 & 1998 Feb 03 16:52 & 6420 & 2086 & 0.077 & orbit & 0.006 & No\tablenotemark{d} \nl
30703-01-08-00 & 1998 Feb 14 23:51 & 4920 & 1195 & 0.054 & orbit & 0.020 & No \nl
\enddata

\tablenotetext{a}{Radio points were interpolated from radio observations within
8 hours of the X-ray observation. Radio fluxes are at a frequency of 8.3 GHz, from public domain data from the NSF-NRAO-NASA Green
Bank Interferometer, unless otherwise indicated.}
\tablenotetext{b}{The QPO here is the one between 0.5-10 Hz with a $Q$ of 
about 10, unless otherwise indicated.}
\tablenotetext{c}{15.2 GHz from the Ryle Telescope, estimated from Figure 4
of Pooley \& Fender (1997).}
\tablenotetext{d}{The QPOs here are either below 0.1 Hz, have $Q$ less than 2,
or both. We do not discuss these QPOs in this paper.}

\end{deluxetable}

\begin{deluxetable}{lccccccccccc}
\scriptsize
\tablecolumns{12}
\tablewidth{0pc}
\tablecaption{PCA Fits\tablenotemark{a} to a Model of Absorption, DISKBB, Power Law or Exponential, and Gaussian\label{fits}}
\tablehead{
\colhead{Date}		& \colhead{DBB} 	
& \colhead{Hard\tablenotemark{c}} & 
\colhead{Hard/} 	& \colhead{$\Gamma$} 	&
\colhead{$E_c$} & \colhead{$N_{\Gamma}$\tablenotemark{d}}  &
\colhead{$T_{col}$} 	& \colhead{$R_{col}$} 	&
\colhead{$E_{gauss}$} & \colhead{$N_{gauss}$} & \colhead{${\chi^2}_{\nu}$} \\
\colhead{UT}		& \colhead{Flux\tablenotemark{b}} &
				\colhead{Flux\tablenotemark{b}} &
\colhead{DBB}		& \colhead{}	& \colhead{keV}	& \colhead{} & 
\colhead{keV}	& \colhead{(km)}	& \colhead{keV} 
& \colhead{$\times 10^{-2}$} 	& \colhead{} 
}
\startdata
1996 May 05 & 2.6 & 2.3 & 0.88 & \nodata & 3.28 & 1.4 & 1.38 & 55 & 6.5 & 3.2 & 1.02 \nl
1996 Jul 11 & 0.2 & 8.4 & 12 & 3.15 & \nodata & 62.1 & 3.87 & 2.1 & 6.0 & 11 & 0.94 \nl
1996 Jul 26 & 0.5 & 3.6 & 2.6 & 2.61 & \nodata & 15.8 & 4.15 & 2.6 & 6.0 & 13 & 0.97 \nl
1996 Aug 03 & 0.4 & 3.8 & 3.0 & 2.69 & \nodata & 18.3 & 4.17 & 2.5 & 6.0 & 12 & 0.90 \nl
1996 Oct 07 & 2 & 2.3 & 1.07  & 2.63 & \nodata & 12.5 & 1.36 & 67 & 6.1 & 3.0 & 1.27 \nl
1996 Nov 28 & 0.4 & 5.8 & 5.3 & 2.63 & \nodata & 26.7 & 1.13 & 33 & 6.1 & 5.5 & 1.16 \nl
1997 Jan 29 & 0.4 & 2.0 & 1.7 & 2.32 & \nodata & 6.3 & 0.87 & 55 & 6.4 & 3.4 & 0.76 \nl
1997 Feb 01 & 0.4 & 1.9 & 1.5 & 2.31 & \nodata & 6.0 & 0.85 & 60 & 6.4 & 3.5 & 0.76 \nl
1997 Feb 09 & 0.3 & 1.7 & 2.1 & 2.15 & \nodata & 4.2 & 0.72 & 67 & 6.3 & 3.9 & 0.83 \nl
1997 Feb 22 & 0.4 & 1.7 & 1.6 & 2.21 & \nodata &4.5 & 0.80 & 61 & 6.4 & 3.4 & 0.80 \nl
1997 Mar 26 & 0.4 & 1.6 & 1.3 & 2.28 & \nodata & 4.8 & 0.82 & 64 & 6.4 & 3.2 & 0.79 \nl
1997 May 18 & 0.7 & 2.6 & 4.0 & 2.42 & \nodata & 9.5  & 1.03 & 57 & 6.4 & 1.9 & 1.05 \nl
1997 May 26 & 0.9 & 3.1 & 3.5 & 2.57 & \nodata & 13.3 & 1.20 & 44 & 6.6 & 1.6 & 1.03 \nl
1997 Jul 20 & 3.4 & 2.0 & 0.59 & \nodata & 3.23 & 1.3 & 1.49 & 54 & 6.3 & 1.7 & 0.77 \nl
1997 Aug 13 & 5 & 4.0 & 0.82 & 2.69 & \nodata & 19.5 & 1.80 & 45 & 6.0 & 2.9 & 1.85 \nl
1997 Aug 14 & 2 & 3.5 & 2.4 & 2.58 & \nodata & 16.3 & 1.13 & 75 & 6.3 & 2.8 & 1.19 \nl
1997 Aug 19 & 5 & 2.8 & 0.64 & 2.87 & \nodata & 16.5 & 1.79 & 43 & 6.0 & 1.0 & 1.96 \nl
1997 Sep 09 & 1 & 3.8 & 3.5 & 2.78 & \nodata & 20.7 & 1.24 & 57 & 6.1 & 2.3 & 1.21 \nl
1997 Sep 16 & 1 & 7.8 & 7.0 & 2.89 & \nodata & 44.1 & 1.52 & 31 & 6.1 & 1.7 & 1.53 \nl
1997 Sep 18 & 0.3 & 11 & 55 & 3.0 & \nodata & 74 & 2.58 & 8 & 6.1 & 7.5 & 2.54 \nl
1997 Oct 09 & 0.3 & 6.7 & 7.5 & 3.12 & \nodata & 48.0 & 4.82 & 1.6 & 6.3 & 8.7 & 0.67 \nl
1997 Oct 25 & 0.3 & 3.7 & 4.5 & 2.66 & \nodata & 17.2 & 4.57 & 1.6 & 6.3 & 7.0 & 1.15 \nl
1997 Oct 30 & 0.9 & 5.0 & 7.0 & 2.63 & \nodata & 23.0 & 1.11 & 55 & 6.4 & 3.0 & 1.16 \nl
1997 Nov 17 & 2.7 & 2.3 & 0.85 & \nodata & 3.42 & 1.3 & 1.49 & 48 & 6.4 & 2.7 & 0.50 \nl
1997 Dec 22 & 5 & 2.9 & 0.58 & 2.68 & \nodata & 14.3 & 1.82 & 44 & 6.0 & 3.4 & 1.87 \nl
1998 Feb 03 & 1.8 & 1.4 & 0.79 & \nodata & 3.24 & 0.9 & 1.30 & 51 & 6.1 & 6.4 & 1.07 \nl
1998 Feb 14 & 1.6 & 0.5 & 0.29 & \nodata & 3.46 & 0.3 & 1.10 & 68 & 6.4 & 7.4 & 2.56 \nl
\enddata
\tablenotetext{a}{All parameters are the averages of the values for each time segment included in the results of Section 3 of this paper. 
We estimate the error introduced by our spectral fitting procedure
by varying an individual parameter (holding the others fixed) 
for each fit until chi-squared for the spectral fits changes by 2.7, which
represents a 90\% confidence interval. 
For PCA spectra collected each satellite orbit or 
every 512 s, the 90\% confidence intervals are typically 
$\Gamma \pm 0.01$, $N_{\Gamma} \pm 0.4$ (for the power law model), $E_c \pm
0.1$, $N_{\Gamma} \pm 0.1$ (for the exponential model), $T_{col} \pm 0.03$ keV, $R_{col}\pm 2$ 
km, $N_{gauss} \pm 0.2\times 10^{-2}$, and $E_{gauss} \pm 0.06$. 
For spectra integrated for 32 s, our uncertainty is typically 
$\Gamma \pm 0.06$, $N_{\Gamma} \pm 4$,  
$T_{col} \pm 0.08$ keV, $R_{col} \pm 3$ km (or 10\%), 
$N_{gauss} \pm 0.5\times 10^{-2}$, and $E_{gauss} \pm 0.2$.}
\tablenotetext{b}{These are unabsorbed values for the flux, calculated as indicated in the text. The units are $10^{-8}$ ergs/cm$^2$sec.}
\tablenotetext{c}{Represents either the flux in power law or exponential, depending on the model used.}
\tablenotetext{d}{Represents either the normalization of the power law or exponential, depending on the model used.}
\end{deluxetable}

\begin{deluxetable}{lccccccccccc}
\scriptsize
\tablecolumns{12}
\tablewidth{0pc}
\tablecaption{PCA/HEXTE Fits\tablenotemark{a} to a Model of Absorption, DISKBB, Cut-Off Power Law or Exponential, and Gaussian\label{hexte}}
\tablehead{
\colhead{Date}		& \colhead{Max E\tablenotemark{b}} & \colhead{DBB}
& \colhead{Hard\tablenotemark{d}} & 
 \colhead{$\Gamma$} 	& \colhead{$E_c$} & 
\colhead{$N_{\Gamma}$}  &
\colhead{$T_{col}$} 	& \colhead{$R_{col}$} 	&
\colhead{$E_{gauss}$} & \colhead{$N_{gauss}$} & \colhead{${\chi^2}_{\nu}$} \\
\colhead{UT} & \colhead{keV}		& \colhead{Flux\tablenotemark{c}} &
		\colhead{Flux\tablenotemark{c}} 
& \colhead{}		& \colhead{keV}  & \colhead{} & 
\colhead{keV}	& \colhead{km}	& \colhead{keV} 
& \colhead{$\times 10^{-2}$} 	& \colhead{} 
}
\startdata
1996 Nov 28 & 150 & 0.6 & 5.8 & 2.49 & 90 & 21.7 & 1.13 & 39 & 6.0 & 4.4 & 1.04 \nl
1997 Jan 29 & 150 & 0.5 & 2.2 & 2.17 & 110 & 4.8 & 0.93 & 52 & 6.7 & 2.7 & 1.03 \nl
1997 Feb 01 & 170 & 0.5 & 2.3 & 2.15 & 100 & 4.5 & 0.90 & 56 & 6.7 & 2.8 & 1.02 \nl
1997 Feb 09 & 170 & 0.3 & 2.0 & 1.94 & 80 & 3.0  & 0.86 & 51 & 6.6 & 3.1 & 1.91 \nl
1997 Feb 22 & 150 & 0.4 & 2.1 & 2.09 & 120 & 3.6 & 0.84 & 55 & 6.7 & 2.7 & 1.08 \nl
1997 Mar 26 & 160 & 0.5 & 2.0 & 2.14 & 110 & 3.8 & 0.86 & 60 & 6.6 & 2.5 & 1.12 \nl
1997 Jul 20 & 50 & 3.5 & 2.0 & \nodata & 3.2 & 1.3 & 1.49 & 55 & 6.3 & 1.9 & 0.65 \nl   
1997 Oct 09 & 90 & 0.2 & 6.1 & 2.97 & \nodata & 37.9 & 4.48 & 1.6 & 6.3 & 7.3 & 1.06 \nl  
1997 Oct 25 & 90 & 0.3 & 4.1 & 2.71 & \nodata & 18.1 & 4.77 & 1.6 & 6.3 & 7.5 & 1.34 \nl
1997 Nov 17 & 40 & 2.8 & 2.3 & \nodata & 3.5 & 1.2 & 1.51 & 47 & 6.4 & 1.8 & 0.71 \nl
1998 Feb 03 & 40 & 1.8 & 1.4 & \nodata & 3.2 & 0.9 & 1.31 & 51 &6.1 & 0.6 & 1.08 \nl
1998 Feb 14 & 30 & 1.6 & 0.5 & \nodata & 3.4 & 0.3 & 1.09 & 70 & 6.3 & 7.8 & 3.17 \nl
\enddata
\tablenotetext{a}{The uncertainties on the parameters from combined PCA/HEXTE 
fits with a cut-off power law are $T_{col} \pm 0.03$ keV, $R_{col} \pm 3$ km,
$\Gamma \pm 0.03$, $E_c \pm 15$, $N_{\Gamma} \pm 0.3$.
When the exponential is used for the hard component, the uncertainties
are typically $T_{col} \pm 0.02$, $R_{col} \pm 1$ km, $E_c \pm 0.1$, and 
$N_{\Gamma} \pm 0.1$.
The errors on the Gaussian are typically $E_{gauss} \pm 0.1$ and $N_{gauss} 
\pm 0.3 \times 10^{-2}$ in all cases.}
\tablenotetext{b}{The maximum energy at which the source dominates 
the background in the energy spectrum was used as the upper limit for the
range the spectral fit was performed on.}
\tablenotetext{c}{These are unabsorbed values for the flux, calculated as indicated in the text. The units are $10^{-8}$ ergs/cm$^2$sec.}
\tablenotetext{d}{Represents either the flux in the power law or exponential, 
depending on the model used.}
\end{deluxetable}

\end{document}